\documentclass[tightenlines,aps,eqsecnum,superscriptaddress]{revtex4}

\usepackage{psfig,amsmath}

\def\alloy{CBr$_4$--C$_2$Cl$_6$}
\def\be{\begin{equation}}
\def\ee{\end{equation}}

\begin{document}

\title{Quantitative phase-field modeling of two-phase solidification}

\author{R. Folch}

\affiliation{Laboratoire de Physique de la Mati\`ere Condens\'ee,
CNRS/\'Ecole Polytechnique, 91128 Palaiseau, France}

\affiliation{Institut-Lorentz, Universiteit Leiden, Postbus 9506, 
2300 RA Leiden, The Netherlands}

\altaffiliation{Present address: Max-Planck-Institut f\"ur Physik
komplexer Systeme, N\"othnitzer Str. 38, 01187 Dresden, Germany}

\author{M. Plapp}

\affiliation{Laboratoire de Physique de la Mati\`ere Condens\'ee,
CNRS/\'Ecole Polytechnique, 91128 Palaiseau, France}

\date{\today}

\begin{abstract}
A phase-field model that allows for quantitative simulations
of low-speed eutectic and peritectic solidification under
typical experimental conditions is developed.
Its cornerstone is a smooth free-energy functional, 
specifically designed so that the stable
solutions that connect any two phases are completely free 
of the third phase.
For the simplest choice for this
functional, the equations of motion for each of the two
solid--liquid interfaces can
be mapped to the standard phase-field model of single-phase 
solidification with its quartic double-well potential. 
By applying the thin-interface asymptotics and 
by extending the antitrapping current previously
developed for this model, all spurious thin-interface corrections 
to the dynamics of the solid--liquid interfaces can be
eliminated. This means that simulation results become 
independent of the thickness $W$ of the diffuse interfaces.
As a consequence, accurate results can be obtained using 
values of $W$ much larger than the physical interface thickness,
which yields a tremendous gain in computational power 
and makes simulations for realistic experimental 
parameters feasible. Convergence of the simulation
outcome with decreasing $W$ is explicitly
demonstrated. Furthermore, the results are compared to a 
boundary-integral formulation of the corresponding free-boundary 
problem. Excellent agreement is found, except in the 
immediate vicinity of bifurcation points, a very sensitive situation
where noticeable differences arise. These differences reveal that,
in contrast to the standard assumptions of the free-boundary problem,
out of equilibrium the diffuse trijunction region of the phase-field model
can (i) slightly deviate from Young's law for the contact angles,
and (ii) advance in a direction that forms a finite angle with
the solid--solid interface at each instant. While the deviation (i) 
extrapolates to zero in the limit of vanishing interface thickness, 
the small angle in (ii) remains roughly constant, which indicates
that it might be a genuine physical effect, present even for an 
atomic-scale interface thickness.
\end{abstract}

\pacs{PACS number(s):}

\maketitle

\section{Introduction}
\label{intro}

The phase-field method has emerged in recent years as a powerful
tool to study interface dynamics in areas such as
solidification \cite{Boettinger02}, solid--solid phase 
transformations \cite{Chen02}, and fluid mechanics \cite{Anderson98,folch99}.
Their common point is that the dynamics of interfaces is coupled to one
or several transport fields, which can lead to the spontaneous emergence
of complex interfacial patterns. The characteristic scale of such 
patterns is typically 
macroscopic, whereas the interfaces are rough on a scale of a few times 
the range of the interatomic forces. Because of this intrinsic scale 
separation, the evolution of pattern-forming interfaces has 
traditionally
been formulated in terms of free-boundary 
problems (FBP), in which the interfaces are assimilated to 
mathematical surfaces without thickness \cite{pelce}.
However, the numerical treatment of FBPs is highly
non-trivial, because one needs to discretize and track 
the sharp boundaries. 

The phase-field method avoids this need
by introducing additional fields to distinguish between phases. 
These ``phase fields'' take different, constant values 
in each bulk phase, and interpolate smoothly between these values
through a diffuse interface of thickness $W$. The equations of motion 
for these auxiliary fields are set up so that one of their intermediate 
level sets, which will represent the interface, obeys the desired FBP
in the so-called ``sharp-interface limit'', in which $W$ tends to $0$.

To construct these evolution equations, 
the physics of diffuse interfaces is typically used as a guide,
and often some qualitative physical meaning can be attached to
the phase field. In solidification, for instance, it can be interpreted as 
an order parameter, and its time evolution considered to be a relaxation 
towards the minimum of a free-energy functional, in the spirit 
of the time-dependent Ginzburg--Landau models for the out-of-equilibrium 
thermodynamics of phase transitions \cite{Hohenberg77}.
The sharp-interface limit of the model is then checked
{\it a posteriori} by matching, order by order, terms 
of asymptotic expansions for the fields in powers of $W$ in a 
region of slow (the bulk) and rapid (the interface) variation of 
the fields. At the lowest order in the interface thickness $W$, 
this procedure is quite straightforward and yields a FBP 
that does not exhibit a dependence on $W$.

However, since the numerical discretization needs to resolve 
the rapid variation of the fields through the interfaces, 
simulations are only feasible using finite $W$. Furthermore,
in order to reduce the gap between the scales of the interface thickness
and the pattern, $W$ often needs to be chosen several
orders of magnitude thicker than in reality.
It is therefore mandatory to assess the influence of  
$W$ on the simulation results. 
This can be done rigorously by going to the next order in $W$ 
in the above matching procedure. In this ``thin-interface limit'',
in which $W$ is small but finite, 
an effective FBP is obtained, which
now generally contains terms that scale with $W$. 

While these terms may reflect genuine (but small) physical effects 
associated with the thickness of the interfaces $W$, they
have to be eliminated either by a judicious choice of the model
and its parameters \cite{kr}, or by additional terms in the
evolution equations tailored to counterbalance these 
effects \cite{folch99}, or by a combination of both \cite{onesided}.
This is necessary to scale up $W$ without altering appreciably 
the outcome of simulations. If all terms in $W$ can be eliminated, 
the outcome is independent of $W$, at least below some value 
of $W$ for which the effective FBP holds (and above which
terms of higher order in $W$ become important). This complete 
elimination was achieved for the first time by Karma and Rappel
for the solidification of a pure substance with equal thermal 
diffusivities in both phases (symmetric model) \cite{kr}.
Their pioneer work made it possible to quantitatively
simulate dendritic growth in three dimensions \cite{3ddendrites}.
More recently, the approach has been extended to the
one-sided model, in which the diffusivity vanishes in the solid
phase, the relevant case for alloy solidification \cite{onesided}.

Industrial alloys often have more than two components and more than
one solid phase.
As a first step to deal with such alloys, we formulate and test here
a phase-field model that carries over the above advances
to {\em two-phase} solidification,
a problem that is both practically relevant in metallurgy and
of interest for the physics of pattern formation:
Eutectic and peritectic two-phase composite structures are, together 
with dendrites, the most common solidification microstructures found 
in industrial alloys \cite{Kurz}. At a binary eutectic or peritectic 
point, two different solid phases coexist with the liquid. 
Under certain conditions, both solids grow simultaneously 
in a cooperative manner (coupled growth). For eutectic alloys,
the resulting composite can consist either of alternate lamellae 
of each solid phase or of rods of one phase immerged in a 
continuous matrix of the other.

Due to the presence of two different solid phases, coupled growth 
can exhibit a rich variety of different patterns. The best studied
setting, used as a relevant example throughout this article, is
the directional solidification of thin eutectic 
samples \cite{Hunt66,Seetha88,Faivre92,Ginibre97,Moulinet01}. 
In this process, samples are pulled from a hot into a cold region
with a constant velocity, and after a transient the solidification 
front reaches a steady state characterized by a fairly uniform 
lamellar spacing. 
This spacing is not intrinsically selected by the system, but depends 
on the growth history. Upon a (large enough) change of pulling velocity,
the front undergoes various instabilities and exhibits many classical
features of nonlinear physics, such as bifurcations to states of lower 
symmetry, periodic limit cycles, spatiotemporal chaos and soliton-like 
traveling perturbations \cite{Faivre92,Ginibre97,Moulinet01,Kassner91,Karma96}.

Numerous phase-field models for two-phase or multi-phase
solidification have been published
\cite{Karma94,Elder94,Wheeler96,Steinbach96,Tiaden98,Nestler00,Lo01,Plapp02,dresden,Kim04}.
While such models can qualitatively \cite{Plapp02,dresden,Kim04}
(or, in some cases, even quantitatively for some aspects \cite{threshold})
reproduce experimental observations, more accurate modeling is 
necessary in order to carry out detailed comparisons to
experiments on specific substances. Albeit all those previous models 
have the correct sharp-interface behavior, it seems quite difficult to
perform a thin-interface analysis of them. 
As a consequence, the influence of the interface thickness
on their results remains uncontrolled.

The reason for this difficulty is intricately linked to the 
construction of the models: The starting point for any asymptotic
analysis is the equilibrium front solution that connects two
different bulk phases. For a binary alloy, there are two or more 
fields involved (the composition and one or more phase fields)
through two or more coupled nonlinear partial differential equations,
generally without known analytic solution. 
In the sharp-interface limit $W\to 0$, the 
coupling between the equations tends to zero, so that a 
separate solution of each equation can be found and used
in the asymptotic analysis; however, in the thin-interface
limit ($W$ small but finite), the coupling terms remain.
A strategy to construct more tractable models is to choose
the coupling terms such that they vanish in equilibrium, 
which obviously facilitates finding
analytic equilibrium solutions for arbitrary coupling strength.
Such a model has recently been developed for a dilute
binary alloy \cite{onesided}. 

For two-phase solidification, an additional difficulty 
arises, because one now needs to distinguish between more than two 
phases. The first phase-field models for eutectic growth 
used the standard solid--liquid phase field, and a 
concentration field \cite{Karma94,Elder94}
or a second phase field \cite{Wheeler96,Lo01}
to distinguish between the two solids, $\alpha$ and $\beta$.
Again, the equilibrium front problem takes the form of at least two
coupled nonlinear partial differential equations with no
known analytic solution.
Later on, it was proposed to assign a separate phase field $p_i$ 
to each phase $i$, indicating its presence ($p_i=1$) or absence 
($p_i=0$). Each phase field can then be understood as a local 
volume fraction, and $\sum_i p_i=1$ locally \cite{Steinbach96}. 
If, on an interface that connects phases $i$ and $j$, no other phase 
is present ($p_k=0$, $k\neq i,j$), the above sum rule permits 
to eliminate one of the two remaining phase fields 
from the evolution equations, and the 
interface can be described by a single variable,
which, again, makes the existence of an exact analytic solution 
more likely. 
However, the stable solution for the phase fields across an
interface {\em does} typically present some amount of the other 
phase(s). Although this recalls the microscopic physical phenomenon 
of adsorption at an interface, this effect has to be eliminated, 
because it would scale with the interface width. Recently, the use 
of the so-called double obstacle potential \cite{2obstacle}  
in the volume fraction formulation \cite{mpfwith2obstacle}
has permitted to reduce or even eliminate the presence of 
supplementary phases in the interfaces. In conjunction with
the so-called interface-field method \cite{interface}, this 
methodology has recently been applied to simulate eutectic 
growth \cite{Kim04}. However, no thin-interface analysis 
of such models is presently available, and might even
be complicated by the singular nature of this free energy.

Having all this in mind, our strategy is to develop a phase-field 
model for two-phase solidification with a {\em smooth}, specifically 
designed free-energy functional that ensures the absence of third 
phases in the interfaces and with a coupling 
between the phase and concentration fields
that vanishes in equilibrium.
Although we use as a starting point
a free-energy functional, we will not seek to link our model 
to explicit thermodynamic alloy models. 
Instead of this, the final goal is 
to simulate the given free-boundary problem (FBP)
as efficiently as possible; the free
energy functional is ``engineered'' in this perspective.

We obtain a model that can accommodate
arbitrary eutectic or peritectic phase diagrams and that exactly
reduces, for each of the two solid--liquid interfaces, 
to the standard phase-field model for the solidification into 
a single solid phase, with the standard quartic double-well 
potential. 
Obviously, this is advantageous, since we can build on the
progress that has recently been made on this model \cite{kr,onesided}.
In particular, we extend to two-phase solidification the
so-called ``antitrapping current'', a phenomenological
addition to the phase-field equations that was recently
introduced in order to achieve a quantitative phase-field
formulation of alloy solidification \cite{onesided}; more
details will be given in Sec.~\ref{1sided}.

We thoroughly test our model in two-dimensional simulations and 
compare the outcome to results obtained with the boundary integral
method \cite{Karma96}. We find that the dynamics of the solid--liquid
interfaces are accurately simulated and independent of the interface
thickness for sufficiently thin interfaces, as predicted by the asymptotic
analysis. However, the dynamics of trijunction points where the three interfaces
meet differs from those traditionally postulated in the classic free-boundary
formulation. These differences, due to the diffuseness of the 
trijunctions, are explored in detail in the numerical simulations
(Sec.~\ref{numerics}).

A preliminary account of some of our results has been given in 
Ref.~\cite{Folch03}; here, we give a detailed presentation of 
both the model and the numerical simulations. The rest of the
paper is structured as follows: We first recall the physics of 
eutectic solidification and write down the classic FBP 
(Sec.~\ref{fbp}). Next, we construct our phase-field
model step by step (Sec.~\ref{pf}); we explore the mapping 
to single-phase solidification
on the solid--liquid interfaces to deduce their
thin-interface behavior, and then refine the model to make this
behavior match as closely as possible the FBP 
of Sec.~\ref{fbp}. In Sec.~\ref{numerics}
we present the numerical simulations and the lessons one can learn
from them. We then conclude with a summary of our main results
and a discussion of further prospects (Sec.~\ref{conclusions}).

\section{free-boundary problem}
\label{fbp}

\begin{figure}
\centerline{\psfig{file=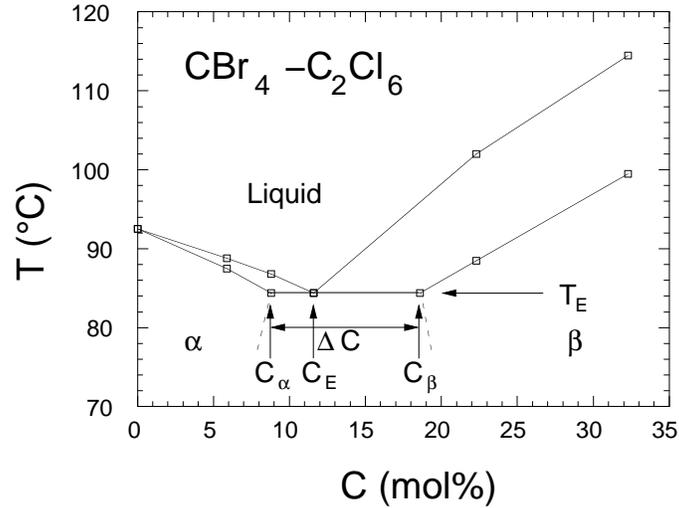,width=.5\textwidth}}
\vskip 1cm
\caption{Experimental phase diagram of the transparent organic eutectic alloy 
CBr$_4$--C$_2$Cl$_6$ (modified from J. Mergy, G. Faivre and S. Akamatsu).
$C$ is the concentration of C$_2$Cl$_6$, and we have indicated the
eutectic temperature $T_E$, the equilibrium concentrations of each phase at
the eutectic temperature, $C_\alpha$, $C_\beta$, and $C_{\rm E}$,
and the width $\Delta C$ of the eutectic plateau.}
\label{phasediag}
\end{figure}

Consider a binary, eutectic alloy. As a relevant example, 
we take the transparent organic mixture 
\alloy, for which accurate and extensive experimental data are 
available \cite{Faivre92,Ginibre97,Moulinet01,mergy}. Its phase diagram
is shown in Fig.~\ref{phasediag}, where $T$ is the temperature and
$C$ the composition, expressed as molar fraction of the solute, 
C$_2$Cl$_6$. The liquidus lines of two solid--liquid equilibria
meet at the eutectic point $(C_{\rm E}, T_{\rm E})$,
the lowest melting point of the alloy. At this temperature,
liquid of composition $C_{\rm E}$ can coexist with two solids of
composition $C_\alpha < C_{\rm E}$ and $C_\beta > C_{\rm E}$. These
compositions define the limits of the eutectic plateau, of total 
width $\Delta C=C_\beta - C_\alpha$.

For slow solidification 
latent heat rejection can be neglected.
The temperature is then constant throughout the system for isothermal
boundary conditions (isothermal solidification). 
For directional solidification in an imposed thermal
gradient, one must further assume equal heat conductivities in all phases
for the temperature field to be independent of the interface position 
and fixed by the experimental setup only. For a
gradient $G$ directed along the $z$ direction with pulling speed
$v_p$ in the negative $z$ direction, we then have
\be
T=T_{\rm E} + G (z-v_p t),
\label{tempsetup}
\ee
where we have chosen the origin of the $z$ axis at the position 
of the eutectic isotherm at $t=0$.

The temperature of a solid--liquid interface is given by the
generalized Gibbs-Thomson condition,
\be
T=T_{\rm E} + m_i (C_+-C_{\rm E})
-\frac{\sigma_{i{\rm L}} T_{\rm E}}{L_i}\kappa - \frac{v_n}{\mu_i}
\label{gibbsthomson}
\ee
where $C_+$, $\kappa$ and $v_n$ are the concentration on the
liquid side of the interface, the local interface curvature,
and its normal velocity, respectively,
and $i=\alpha,\beta$. 
Up to the first two terms on its r.h.s., Equation (\ref{gibbsthomson})
is the equation of the liquidus lines, where $m_i$ are the liquidus 
slopes ($m_\alpha<0$, $m_\beta>0$), taken at the eutectic point, 
but assumed to be independent of the concentration.
The third term on the r.h.s, where $\sigma_{i{\rm L}}$ are the solid--liquid surface 
tensions and $L_i$ the latent heats of fusion per unit volume,
represents the capillary effect which lowers the melting temperature
of convex parts of the solid. Finally, the fourth term stands for
the effects of interface kinetics; $\mu_i$ therein is
the linear kinetic coefficient, defined as the 
proportionality constant between the local undercooling and 
the velocity of a flat interface. 
Eliminating the
temperature between Eqs.~(\ref{tempsetup}) and (\ref{gibbsthomson}),
we obtain boundary conditions for the concentrations at an
interface of specified position, velocity, and curvature 
[Eq. (\ref{githo}) below].

Growth is limited by solute transport, much slower than that of heat.
For solidification in thin samples, convection in the liquid is
suppressed, and solute transport occurs by diffusion only.
Furthermore, diffusion in the solid is in most cases so slow,
that the final distribution of impurities is still
the trace left by the concentration at the solid
side of the solidification front $C_-$, which is 
assumed to follow the equilibrium partition relation
\be
\label{partition}
C_- = k_i C_+,
\ee
where $k_i$ are the partition coefficients.
Therefore, we neglect diffusion in the solid phases (one-sided model).
Then, the velocity of the interface and the diffusion flux
on the liquid side are related by the Stefan condition,
\be
v_n (C_+-C_-) = - \hat n \cdot (D \vec\nabla C|_+),
\ee
where $\hat n$ is the unit vector normal to the interface and
pointing into the liquid. This condition expresses mass
conservation: 
The concentration difference between the 
two phases has to be transported away by diffusion for the interface 
to move. As a corollary, in the one-sided model the
solid--solid interfaces cannot move, and hence their
shape is just the trajectory followed by the trijunctions.

All of the above already specifies the dynamics of each
solid--liquid interface independently. Each of them is thus
found to obey the following free-boundary problem:  
\begin{subequations}
\label{fbpeqs}
\begin{eqnarray}
\label{diffusion}
\partial_t c & = & D \nabla^2 c\\
\label{mc}
D \hat n \cdot \vec\nabla c|_+ & = & [c_i - (1-k_i) c_+] v_n\\
\label{githo}
c_+ & = & \mp \left(\frac{z_{\rm int}-v_p t}{l_T^i} + 
   d_i \kappa + \beta_i v_n\right),
\end{eqnarray}
\end{subequations}
where we have defined the scaled concentration field
\be
\label{scaledcfield}
c = \frac{C-C_{\rm E}}{\Delta C}
\ee
and the scaled equilibrium concentrations of the two solids
at the eutectic temperature,
\be
\label{scaledcsolids}
c_i = \frac{C_i - C_{\rm E}}{\Delta C}.
\ee
Equation (\ref{diffusion}) holds in the liquid, and Equations (\ref{mc})
and (\ref{githo}) on
the interface; 
the minus (plus) sign in Eq.~(\ref{githo}) is valid for
the $\alpha$--liquid ($\beta$--liquid) interface, $z_{\rm int}$
is the $z$ position of the interface,
\be
\label{thermaldef}
l_T^i = \frac{|m_i|\Delta C}{G}
\ee
are the thermal lengths,
\be
\label{capillarydef}
d_i = \frac{\sigma_{i{\rm L}} T_{\rm E}}{L_i |m_i| \Delta C},
\ee
the capillary lengths, and
\be
\beta_i = \frac{1}{\mu_i |m_i| \Delta C},
\ee
kinetic coefficients. We also define the average capillary
and thermal lengths, $\bar d = (d_\alpha + d_\beta)/2$
and $\bar l_T = (l_T^\alpha+l_T^\beta)/2$, as well as
the diffusion length
\be
l_D = D/v_p.
\ee

\begin{figure}
\centerline{\psfig{file=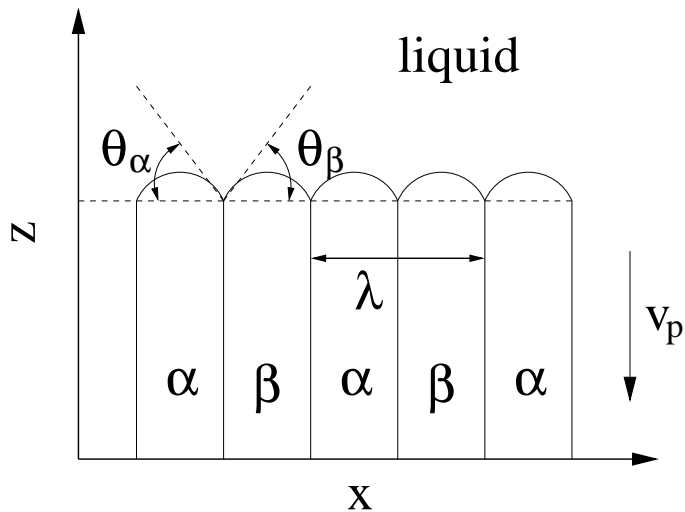,width=.5\textwidth}}
\caption{Sketch of the lamellar geometry with the definitions
of the contact angles $\theta_\alpha$ and $\theta_\beta$;
$\eta$ is the volume fraction of $\alpha$ phase, and $\lambda$ 
the lamellar spacing.}
\label{figeutstat}
\end{figure}

Finally, the $\alpha$--liquid and $\beta$--liquid interfaces are connected
by imposing local equilibrium at trijunction points where three
interfaces ($\alpha$--liquid, $\beta$--liquid, and $\alpha$--$\beta$) 
meet. The balance of the surface tension at the trijunction
points reads
\begin{equation}
\label{young}
\hat t_{\alpha{\rm L}}\sigma_{\alpha{\rm L}} +
\hat t_{\beta{\rm L}}\sigma_{\beta{\rm L}} +
\hat t_{\alpha\beta}\sigma_{\alpha\beta} = \vec 0,
\end{equation}
where $\hat t_{ij}$ is a unit vector tangential to the $i$--$j$ 
interface and pointing outwards from the trijunction. This 
condition, known as Young's law, yields the equilibrium
angles between any two interfaces. An important
special case is a steady-state composite structure such
as the lamellae depicted in Fig.~\ref{figeutstat}, for
which the solid--solid interfaces are aligned with the
$z$ direction. The angles of the $i$--liquid interfaces
with the $x$ direction, called contact angles (we adopt
the notation of Refs. \cite{Langer80,Datye81}), are
given by the solution of the equations
\begin{eqnarray}
\sigma_{\alpha\beta} &=& \sigma_{\alpha\rm L} \sin \theta_\alpha + 
                          \sigma_{\beta\rm L} \sin \theta_\beta  \\
\sigma_{\alpha\rm L} \cos\theta_\alpha & = & \sigma_{\beta\rm L} \cos\theta_\beta.
\end{eqnarray}
For equal solid--liquid surface tensions, 
we have $\sin\theta_\alpha = \sin\theta_\beta = 
\sigma_{\alpha\beta}/(2\sigma_{i\rm L})$.

Note that we have not taken into account crystallographic effects,
which would appear through anisotropies in the kinetic coefficients
and the surface tensions; the latter would also lead to additional
``Herring torque'' terms in Young's law. It has been shown by
boundary integral simulations \cite{Kassner91,Karma96} that
all the relevant instabilities and morphologies can be reproduced
by the above model. Therefore, we have limited ourselves to an
isotropic formulation both for the FBP and the phase-field model.

The growth of lamellar composites has been analyzed in the
seminal paper by Jackson and Hunt \cite{jh}. They found that there
exists a family of steady-state solutions parameterized by
the lamellar spacing $\lambda$ (see Fig.~\ref{figeutstat}).
The average undercooling of the front, $\Delta T$, depends
on the spacing as
\be
\label{jhglobal}
\Delta T = \frac{\Delta T_{{\rm JH}}}{2}
  \left(\frac{\lambda}{\lambda_{{\rm JH}}}+\frac{\lambda_{{\rm JH}}}{\lambda}\right),
\ee
where
\begin{eqnarray}
\label{jhdeltat}
\Delta T_{{\rm JH}} & = & \frac{2\sqrt{2} \Delta C}{\eta(1-\eta)}
    \left(\frac{m_\alpha m_\beta}{m_\alpha+m_\beta}\right)
      \sqrt{P(\eta)[(1-\eta)d_\alpha \sin\theta_\alpha+ \eta d_\beta \sin\theta_\beta]/l_D} \\
\label{jhlambda}
\lambda_{{\rm JH}} & = & \sqrt{2 l_D 
  [(1-\eta)d_\alpha \sin\theta_\alpha+ \eta d_\beta \sin\theta_\beta]/P(\eta)},
\end{eqnarray}
with $P(\eta)=\sum_{n=1}^\infty \sin^2(\pi\eta n)/(\pi n)^3$, and
$\eta$ the nominal volume fraction of the $\alpha$ phase at the
eutectic temperature, related to the global sample composition
$c_\infty$ by $c_\infty = \eta c_\alpha + (1-\eta) c_\beta$ \cite{Langer80,Datye81,jh}.
The front undercooling has a minimum ($\Delta T=\Delta T_{{\rm JH}}$)
for the spacing  $\lambda_{{\rm JH}}$, which constitutes a reference 
length for lamellar eutectics.
%

\section{Phase-field formulation}
\label{pf}


In this section we construct and justify our phase-field model.
We start by introducing our notation and a variational formulation 
(Sec.~\ref{framework}).
Next, we introduce a free-energy functional 
that provides the minimal features of the model, i.e.,
three distinct phases with interfaces that are free of any 
adsorbed third phase, and enough freedom to fit a
phase diagram (Sec.~\ref{minimalmodel});
The equilibrium properties of the model, including the phase diagram,
are then verified in Sec. \ref{equilibrium}.
The model is analyzed for a two-phase system, and
it is found that it exactly reduces to the usual phase-field
model of single-phase solidification (Sec.~\ref{mapping}). 
This mapping allows
us to deduce the thin-interface behavior of two-phase 
interfaces without performing a detailed asymptotic analysis.
Furthermore, this analogy serves us as a guideline to
extend the model, and, in particular, to allow for two
independent kinetic coefficients on the two solid--liquid
interfaces (Sec.~\ref{indkinetics}), to introduce a
non-variational formulation (Sec.~\ref{nonvariational}),
to address the one-sided case in which the solid
diffusivity is neglected (Sec.~\ref{1sided}),
and the case of different surface tensions (Sec.~\ref{unequal}).

Although we use as a starting point
a free-energy functional, we should stress that 
we will not seek to link our
model to an explicit thermodynamic formulation. The final
goal is to obtain a model that simulates the free
boundary problem of the previous section as efficiently as possible; 
the free-energy functional is ``engineered'' in this perspective.

\subsection{Framework and notation}
\label{framework}

We use three phase fields $p_i$, $i=\alpha,\beta,{\rm L}$,
and denote $\vec p\equiv (p_\alpha,p_\beta,p_{\rm L})$. Each 
field represents the volume fraction of a different phase,
and hence we would like $p_i\in [0,1]\,\forall i$ and
\begin{equation}
\label{constraint}
\sum_i p_i=1.
\end{equation}
We then introduce a free-energy functional 
\be
\label{fdimensional}
F = \int_V f dV,
\ee
defined as the volume integral of a free-energy density
\be
\label{fedensity}
f(\vec p, \vec\nabla \vec p, c,T)= 
K f_{\rm grad}(\vec\nabla \vec p) 
+ H f_p(\vec p) 
+ X f_c(\vec p,c,T),
\ee
whose dependencies 
have been broken down in the following way:
$f_p$ depends only on the phase fields and provides a
``free-energy landscape'' in $\vec p$; 
in particular, it contains 
the analog of the double-well potential 
of single-phase solidification.
$f_c$ couples the phase fields
to concentration and temperature and hence defines the phase diagram.
It does not contain a term in $(\vec\nabla c)^2$,
which turns out to be unnecessary.
Both $f_p$ and $f_c$ are dimensionless;
$H$ and $X$ are constants with dimensions of energy 
per unit volume.
$f_{\rm grad}$ sets a free-energy cost for gradients in $\vec p$,
forcing interfaces to have a finite width. We will use a
quadratic form in the gradients of the phase fields, and thus
generalize the squared-gradient term of single-phase 
solidification. Therefore, $K$ has dimensions of energy per unit length.

In a variational formulation of the equations of motion,
the non-conserved phase fields are assumed to evolve 
towards the minimum of $F$,
\begin{equation}
\label{modelforp}
\tau(\vec p) \frac{\partial p_i}{\partial t}= 
-\frac{1}{H}
\left.\frac{\delta F}
{\delta p_i}\right|_{p_\alpha + p_\beta + p_{\rm L} = 1} 
  \;\forall i,
\end{equation}
where $H$ has been introduced on the right hand side to remove 
the dimensions of $F$, and $\tau(\vec p)$ is a relaxation time 
that is the same in all Equations (\ref{modelforp}) (i.e., $\forall i$), 
but may depend on the local values of the phase fields
(see Appendix \ref{timeconstants} for a 
detailed motivation of this choice).
The functional derivative has to be evaluated subject to the
constraint of Eq.~(\ref{constraint}). This can be done by 
the method of Lagrange multipliers: A term $\xi (1-\sum_j p_j)$ 
is added to the free-energy functional; the derivatives are 
taken as if the variables $p_i$ were independent, and $\xi$ 
is determined and eliminated. The result for three phases is
\begin{equation}
\label{prescription}
\left. \frac{\delta F}{\delta p_i}\right|_{p_\alpha + p_\beta + p_{\rm L}= 1} =
\frac{\delta F}{\delta p_i} - \frac{1}{3}\sum_j \frac{\delta F}{\delta p_j},
\end{equation}
where the variations on the r.h.s. are now taken as 
if all $p_i$ were independent. We recall that variations 
are evaluated according to
\begin{equation}
\label{funcderiv}
\frac{\delta F}{\delta p_i(\vec x)} = 
\frac{\partial f}{\partial p_i} - \sum_\nu \partial_\nu 
   \frac{\partial f}{\partial (\partial_\nu p_i)},
\end{equation}
where $\nu$ counts the spatial coordinates. Equations~(\ref{modelforp})
and (\ref{prescription}) automatically ensure
$\sum_i \partial p_i/\partial t=0$, which is necessary to 
maintain consistency with the constraint of Eq.~(\ref{constraint}).
Note that we do not impose at this level $p_i\in [0,1]\,\forall i$,
in contrast to other formulations \cite{mpfwith2obstacle}.
This restriction will instead result from the
specific design of our free-energy functional.

The concentration $c$ is a conserved field, and thus obeys the 
continuity equation
\begin{equation}
\label{modelforc}
\frac{\partial c}{\partial t} + \vec\nabla\cdot\vec J = 0,
\end{equation}
where $\vec J$ is the flux of scaled concentration $c$,
\be
\label{dimensionalcurrent}
{\vec J}=-M(\vec p)\vec\nabla \frac{\delta F}{\delta c},
\ee
with $M(\vec p)$ a mobility.

As already mentioned in the introduction, the key point of our 
model is that an interface between any two phases $i$ and $j$ 
should be free of any adsorbed third
phase, that is, $p_k\equiv 0 \;\; \forall k\neq i,j$. 
If we construct a free-energy functional ${\cal F}$ such that 
$p_k= 0,1$ be the only homogeneous, stationary stable solutions 
of Eq. (\ref{modelforp}) for $p_k$,
then a $i$--$j$ interface will be free of phase $k$,
since any perturbation around this solution would relax 
back to zero. More generally, this requirement will ensure that
bulk phases $p_k=1$ are stable, $i$--$j$ interfaces are
stabilized only by the presence of bulk phases $i$ and $j$ at 
each side and contain no third phase, triple junctions or
lines can only exist where interfaces meet, and so on.
Note also that, with appropriate initial and boundary conditions,
this guarantees that $p_i\in[0,1]\,\forall i$ (for the continuum
equations; in the discretized equations, small overshoots may 
occasionally appear). The requirements on ${\cal F}$ for 
stationarity and stability read respectively
\begin{subequations}
\label{valleys}
\begin{eqnarray}
\label{flatness}
\left. \frac{\delta F}{\delta p_k} 
\right |_{p_\alpha + p_\beta + p_{\rm L} = 1,\;\;p_k=0,1} & = & 0 \;\;\forall k, \\
\label{convexity}
\left. \frac{\delta^2 F}{\delta p_k^2} 
\right |_{p_\alpha + p_\beta + p_{\rm L} = 1,\;\;p_k=0,1} & > & 0 \;\;\forall k.
\end{eqnarray}
\end{subequations}
Furthermore, in order to ensure stability with respect
to concentration fluctuations, we need to require
\be
\label{musecond}
\frac{\delta^2 F}{\delta c^2} > 0,
\ee
which is just a standard condition of thermodynamic stability.

To construct a suitable free-energy functional, our strategy
is to choose the two parts of the free-energy density that 
depend on the gradients only ($K f_{\rm grad}$) and the
local values of the fields only ($H f_p + X f_c$), respectively, 
such that the volume integral of each one separately satisfies 
the above conditions.

\subsection{Minimal model}
\label{minimalmodel}

For clarity of exposition, we first construct our model taking
the parts of the free-energy density that functionally depend on the
phase fields only,
$f_{\rm grad}(\vec\nabla p)$ and $f_p(\vec p)$, 
to be invariant under the exchange of any two phases.
We will later break this symmetry in different ways as it
becomes necessary.

For the energetic cost of the gradients of the phase fields, 
$f_{\rm grad}$, we adopt the simplest scalar expression
that is a regular and isotropic function of $\vec\nabla \vec p$
and respects this symmetry,
\begin{equation}
\label{fgrad}
f_{\rm grad}=\frac{1}{2}\sum_i \left(\vec\nabla p_i\right)^2.
\end{equation}
It is straightforward to check that the volume integral of
this function satisfies Eqs. (\ref{valleys}). 

As for the potential part $f_p$, we take
\be
f_p = f_{\rm TW},
\ee
where the triple-well potential $f_{\rm TW}$ is the analog of the 
double well in the standard phase-field model of solidification
with a single solid phase. 
In order to acquire 
some ``geometric intuition'' about its construction, it is useful
to visualize it with the help of the Gibbs simplex. We recall (see 
also Appendix \ref{timeconstants}) that the Gibbs simplex is an equilateral triangle
of unit height, where each vertex represents a different phase 
(either $\alpha$ or $\beta$ solid, or liquid), 
and, for any point, the (signed) distance to the side opposite
to a given vertex represents the value of the phase field 
associated to that vertex. Thus, each side corresponds to a purely 
binary interface, and the center, $p_\alpha=p_\beta=p_{\rm L}=1/3$, 
to the triple point. Points outside the triangle represent 
negative values of one or two phase fields.

The parts of the free-energy density that depend only on the local
values of the fields, but not on their gradients, $f_p$ and $f_c$,
can be plotted as a surface over this simplex. Moreover, the conditions
of Eqs.~(\ref{valleys}) on their volume integrals
can be rewritten using Eq.~(\ref{funcderiv}) 
to replace functional with partial
derivatives: Focusing on the part that we are now concerned with,
$f_p=f_{\rm grad}$, the conditions
read $\partial f_p/\partial p_k|_{p_\alpha + p_\beta + p_{\rm L}= 1,\; p_k=0,1}=0$
and 
$\partial^2 f_p/\partial p_k^2|_{p_\alpha + p_\beta + p_{\rm L}= 1,\; p_k=0,1}>0$.
It can be easily verified that the left hand side of 
the first condition is equal to the slope of the $f_p$ surface, 
plotted over the Gibbs simplex, in the direction $k$.
Therefore, the conditions 
of Eqs. (\ref{flatness}) and (\ref{convexity}) can be understood 
as requiring zero slope (flatness) and convexity, respectively, 
in the direction normal to a side, both on that side
and on the vertex opposite to it. In other words, the sides of 
the triangle must be valleys of the free-energy landscape,
so that the stable solutions that connect two vertices 
run exactly along the sides (purely binary interfaces), and each vertex 
should be a local minimum; hence, there will necessarily 
be a saddle point on each side that will connect the two vertices of the side, 
and each interface will thus ``feel'' a double-well 
potential. There should be no other minimum.

\begin{figure}
\centerline{\psfig{file=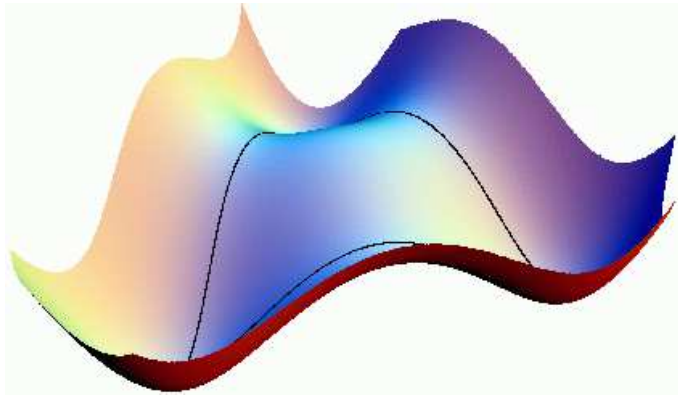,width=.5\textwidth}}
\vskip 1cm
\caption{Triple well potential, visualized as a surface over 
the Gibbs simplex. The black lines mark the borders of the
Gibbs simplex, drawn on the surface; for this function, 
they also coincide with the
trajectory of the equilibrium interface solutions.}
\label{3wellplot}
\end{figure}

The requirement for minima at $p_k=0,1$ along the $p_k$ direction
[Eqs. (\ref{valleys})] suggests
to construct the triple-well potential as a sum
of equal double-well potentials $f_{\rm DW}(p)$ for all
the phase fields: 
\begin{equation}
\label{triplewell}
f_{\rm TW}=\sum_i f_{\rm DW}(p_i).
\end{equation}
If $f_{\rm DW}(p)$ has two equal minima at $p=0,1$, $f_{\rm TW}$ 
will have equal global minima in each phase $i$ characterized
by $p_j=\delta_{ij}$. Applying the prescription of 
Eq.~(\ref{prescription}) for $p_k=0,1$, we find
that the slope is equal to $(2/3)(df_{\rm DW}/dp)(p=p_k) 
- (1/3)(df_{\rm DW}/dp)(p=p_i) - (1/3)(df_{\rm DW}/dp)(p=p_j)$. 
Since $p_k=0,1$ and $(df_{\rm DW}/dp)(p=0,1)=0$, the first term is zero;
furthermore, at $p_k=1$ $p_i=p_j=0$ and the other two terms vanish too;
as for $p_k=0$, then $p_j=1-p_i$, and it can be seen
that flatness is satisfied whenever $f_{\rm DW}(1-p)=f_{\rm DW}(p)$, 
i.e., if $f_{\rm DW}$ is even in $2p-1$. Similarly, it can be 
shown that convexity is also satisfied as long as the double 
well convexity in the wells overcomes its concavity 
on the maximum in between, $2(d^2f_{\rm DW}/dp^2)(p=0,1) > 
-(d^2f_{\rm DW}/dp^2)(p=1/2)$.
The standard quartic double well, proportional to 
\begin{equation}
\label{2well}
f_{\rm DW} = p^2 (1-p)^2,
\end{equation}
satisfies the above requirements. This choice is particularly
convenient since it will allow us to relate our model to the 
standard phase-field model. The result for $f_{\rm TW}$ is plotted 
in Fig. \ref{3wellplot}.
%
%

So far, all bulk phases correspond to equally deep wells, which means
that they all have the same bulk free-energy density. However,  
the relative stability of phases generally depends on the local values 
of the concentration and the temperature. In the phase-field model
for single-phase solidification, this is implemented by adding a
function to the free-energy density that tilts the double well
by an amount that is proportional to the local driving force
for phase transformations. Here, we need to construct a function 
that tilts the triple well in such a way that the conditions of 
flatness (and, if possible, convexity), are maintained.
Note that the condition of flatness in particular, Eq. (\ref{flatness}), 
not only ensures the absence of third-phase adsorption, but also that
the two wells of every pure $i$--$j$ interface remain at $p_i=0,1$ in
spite of the tilt, a condition known to be important for single-phase
solidification \cite{kr}.

We begin by constructing a function $g_i$ which raises the 
well $i$, $g_i(p_i=1)=1$, and leaves the other two as well as the
entire interface between them unaffected, $g_i(p_i=0)=0$.  
On the other two interfaces ($p_j=0$ with $j\neq i$), 
we require $g_i$ to be antisymmetric with respect to the saddle 
point of $f_{\rm TW}$ at $p_i=p_j=1/2$, i.e.,
\be
\label{gsymmetry0}
g_i(1-p_i,p_i,0) = 1 - g_i(p_i,1-p_i,0).
\ee
This requirement is necessary to avoid undesirable thin-interface
corrections in the matching to the free-boundary problem
(see Refs. \cite{onesided,almgren})
and to adjust surface tensions independently of the phase diagram
(see next subsection), and is hence important \cite{foot3}.
Replacing $1-p_i$ by $p_j$ and then exchanging 
the mute indices $i$ and $j$, 
we can rewrite the requirement in the form
\be
\label{gsymmetry}
g_j(p_i,p_j,0) = 1 - g_i(p_i,p_j,0),
\ee
which can be understood as the analog of $p_j=1-p_i$ for $p_k=0$.
This latter form [Eq. (\ref{gsymmetry})] will be used below.
We also impose zero slope on all sides of the Gibbs simplex in
accordance with Eq. (\ref{flatness}),
but not convexity, for reasons which will soon become apparent.  

\begin{figure}
\centerline{\psfig{file=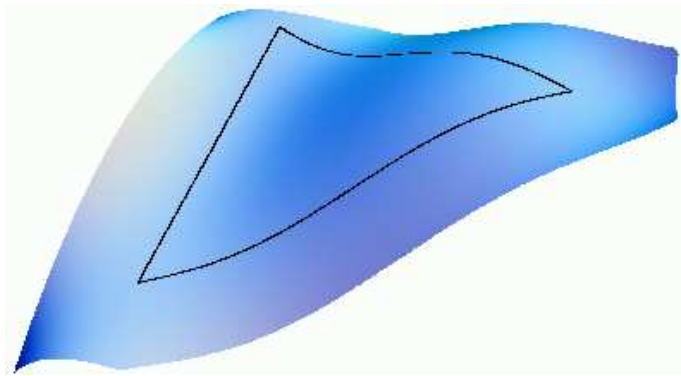,width=.5\textwidth}}
\vskip 1cm
\caption{Elementary tilting function $g_i$, visualized over 
the Gibbs simplex. The $i$ vertex is to the right.}
\label{tiltplot}
\end{figure}

The lowest-order polynomial in the
phase fields that satisfies all the above requirements
is fifth-order and unique at this order, 
\begin{equation}
\label{tilt}
g_i=\frac{p_i^2}{4} \left \{15(1-p_i)\left [1+p_i-(p_k-p_j)^2\right ]
+ p_i \left (9p_i^2-5\right )\right \};
\end{equation}
it is plotted in Fig. \ref{tiltplot}.
Remarkably enough, this function reduces to the polynomial
$p_i^3(10-15p_i+6p_i^2)$ on the $i$--$j$ and $i$--$k$ interfaces
when the constraint $p_i+p_j+p_k=1$ is taken into account, 
which happens to be proportional to the tilting function used 
in the available quantitative models of single-phase 
solidification \cite{kr,onesided}.

%

We then couple the phase fields to the concentration $c$ 
and the temperature $T$ through the functions $g_i(\vec p)$: 
We use these functions to interpolate
between three free-energy
parabolic wells in concentration 
\be
\label{fci}
f_{c,i}=(c-A_i)^2/2 + B_i
\ee
for the three phases $i=\alpha,\beta,\rm L$, characterized by
$p_j=g_j=\delta_{ij}$:
\begin{equation}
\label{fc}
f_c=\frac{1}{2}\left [c-\sum_i A_i(T) g_i(\vec p)\right ]^2 
+ \sum_i B_i(T) g_i(\vec p).
\end{equation}
The $A_i(T)$ and $B_i(T)$ define the
equilibrium phase diagram, as discussed in the next subsection.

With $f_c$ specified, we can now write down an explicit
expression for the concentration current. We define a 
dimensionless chemical potential,
\be
\label{muexplicit}
\mu=\frac{1}{X}\frac{\delta F}{\delta c}=\frac{\partial f_c}{\partial c}= 
c-\sum_i A_i(T) g_i(\vec p),
\ee
so that Eq.~(\ref{dimensionalcurrent}) becomes
\begin{equation}
\label{modelforcurrent}
\vec J = -D(\vec p)\vec\nabla\mu,
\end{equation}
where $D(\vec p)=M(\vec p)X$ is a phase-dependent diffusivity.
Note that the $f_c$ in Eq. (\ref{fc}) or (\ref{muexplicit})
satisfies the stability condition of Eq.~(\ref{musecond}).

Equation (\ref{flatness}) for no third-phase absorption also 
holds for $f_c$, since each of the $g_i$ satisfies this condition 
by construction. However, no general statement can be made about 
the convexity of $f_c$, because it depends on the form of $g_i$
and on the values of $A_i$, $B_i$ and c. It can be shown that for
special choices of the $A_i$ and $B_i$, the second derivatives 
of $f_c$ with respect to the third phase can be made to vanish 
throughout all borders of the Gibbs simplex. In any case,
the other components of the free-energy density will ensure
global convexity and hence stability for sufficiently small
values of the ratio $X/H$. The solutions
can always be stabilized by adding ``obstacle'' terms in the free energy, as 
discussed in Sec. \ref{unequal} below; however, our experience is that, 
whenever an instability occurs, the model is anyway too far from the 
thin-interface limit to represent the correct free-boundary problem,
so that the convexity of $f_c$ is not an issue in practice.

\subsection{Equilibrium solutions}
\label{equilibrium}

In order to illustrate how the evolution equations of the model 
are derived and to clarify the reasons for 
the particular choice of Eq.~(\ref{fc}) for $f_c$, 
we derive the stationary solution for a planar interface. 
We check that this solution
satisfies the conditions of thermodynamic equilibrium,
compute its surface tension, and obtain the relationship
between the coefficients $A_i(T)$ and $B_i(T)$ and
the equilibrium phase diagram.

\subsubsection{Evolution equations}
We first write down explicitly the evolution equations for the 
minimal model, whose building blocks have been derived above. 
For the phase fields, 
we take the functional derivative of the free energy
with respect to each phase field $p_i$ according to Eq. (\ref{prescription}),
where the component of the free-energy density $f_{\rm grad}$ is given by
Eq.~(\ref{fgrad}), $f_p=f_{TW}$ by Eq.~(\ref{triplewell}) with Eq.~(\ref{2well}), 
and $f_c$ by Eq.~(\ref{fc}) with Eq.~(\ref{tilt}). 
Since $c$ and $\mu$ are related, we can
eliminate one in order to reduce the number of variables;
we choose to eliminate $c$. We obtain:
\begin{eqnarray}
\label{evolutionofpmin}
\tau(\vec p)\frac{\partial p_i}{\partial t} & = &
W^2\nabla^2p_i
+ \frac{2}{3}\left [ -2p_i(1-p_i)(1-2p_i) + \sum_{j\neq i} p_j(1-p_j)(1-2p_j) \right ] 
   \nonumber \\
& & \qquad\mbox{}
\label{fcderivativemin}
+ \tilde\lambda\sum_j 
\left.\frac{\partial g_j}{\partial p_i}\right |_{p_\alpha+p_\beta+p_{\rm L}=1} 
\left ( \mu A_j - B_j \right )
\quad \forall i, \\
\label{evolutionofmumin}
\frac{\partial \mu}{\partial t} & = &
\vec\nabla\cdot \left [D(\vec p)\vec\nabla\mu\right ]
-\sum_i A_i \frac{\partial g_i(\vec p)}{\partial t},
\end{eqnarray}
where we have defined
$W=\sqrt{K/H}$, $\tilde\lambda =X/H$, and
\be
\label{gipi}
\left.\frac{\partial g_i}{\partial p_i}\right |_{p_\alpha+p_\beta+p_{\rm L}=1}
= \frac{5}{2} p_i \left[ (p_k-p_j)^2 (3p_i-2) + (1-p_i)^2 (3p_i+2) \right ],
\quad i\neq j,k, \quad j\neq k,
\ee
\be
\label{gjpi}
\left.\frac{\partial g_j}{\partial p_i}\right |_{p_\alpha+p_\beta+p_{\rm L}=1}
= -\frac{1}{2} 
\left.\frac{\partial g_i}{\partial p_i}\right |_{p_\alpha+p_\beta+p_{\rm L}=1}
 + \frac{15}{2} p_j^2 (1-p_j) (p_k-p_i),
\quad j\neq i, \quad k\neq i,j.
\ee
The evolution equation for $\mu$ is a diffusion 
equation with a source term that reflects the redistribution of solute 
at advancing interfaces.

Next, consider a generic $i$--$j$ interface. 
Since we have taken care that our free-energy functional,
if not all of its components, satisfies 
Eqs. (\ref{valleys}) for $p_k=0,1$ ($k\neq i,j$),
the phase field that corresponds to the third phase $k$ vanishes, 
$p_k=0$. Therefore, the sum rule of 
Eq.~(\ref{constraint}) can be used to eliminate $p_j$ through $p_j=1-p_i$
[and hence $g_j$ in terms of $g_i$, Eq. (\ref{gsymmetry}), $g_k(p_k=0)=0$], 
so that we are left with a single phase field $p_i$ as desired.
Equations~(\ref{evolutionofpmin}--\ref{evolutionofmumin}) thus 
yield \cite{footonehalf}
\begin{eqnarray}
\tau(\vec p) \frac{\partial p_i}{\partial t} & = &
  W^2 \nabla^2 p_i - 2p_i(1-p_i)(1-2p_i) \nonumber \\
&& \qquad\mbox{} - 15\tilde\lambda p_i^2(1-p_i)^2 
  \left [ \mu\left (A_j-A_i\right) - \left (B_j-B_i\right) \right ]
\label{ponij} \\
\label{muonij}
\frac{\partial\mu}{\partial t} & = &
\vec\nabla\cdot \left [D(\vec p)\vec\nabla\mu\right ]
+ (A_j-A_i) 
  \frac{\partial g_i(p_k=0)}{\partial t},
\end{eqnarray}
where $g_i(p_k=0)$ is evaluated using $p_j=1-p_i$. 

\subsubsection{Stationary solutions}
We search for stationary planar interface solutions of 
Eqs.~(\ref{ponij}) and (\ref{muonij}), that is, solutions that
have vanishing time derivatives but vary in space along a
single coordinate $x$ and connect bulk phases $i$ [$p_i(x=-\infty)=1$] 
and $j$ [$p_i(x=+\infty)=0$]. From the condition $\partial \mu/\partial t=0$
we obtain that the chemical potential $\mu$, if bounded, has to be 
a constant in space. This is of course one of the conditions of 
thermodynamic equilibrium, and we hence denote this constant by 
$\mu_{\rm eq}^{ij}$. Its value is fixed by the second requirement,
$\partial p_i/\partial t=0$, through a solvability condition for
the stationary, one-dimensional version of Eq. (\ref{ponij}),
\be
  W^2 \partial_{xx} p_i = 2p_i(1-p_i)(1-2p_i) + 15\tilde\lambda p_i^2(1-p_i)^2 
  \left [ \mu\left (A_j-A_i\right) - \left (B_j-B_i\right) \right ] = 
  - \frac{d V(p_i)}{dp_i},
\label{mechanical}
\ee
where the second equality defines a ``potential'' 
$V(p_i)=-p_i^2(1-p_i)^2-(\tilde\lambda/2) p_i^3(10-15p_i+6p_i^2) 
  \left [ \mu\left (A_j-A_i\right) - \left (B_j-B_i\right) \right ]$
up to a constant. 
This notation refers to the well-known ``particle-on-a-hill'' 
analogy, in which Eq.~(\ref{mechanical}) is interpreted as an
equation of motion with time coordinate $x$ for a point particle
of position $p_i$ that moves in the potential $V$. Since there is
no dissipative term, a solution to this equation that connects
$p_i=0$ and $p_i=1$ exists if and only if $V(0)=V(1)$,
which requires the squared bracket
$[ \mu\left (A_j-A_i\right) - \left (B_j-B_i\right) ]$
to vanish and hence yields
\begin{equation}
\label{mueq}
\mu_{\rm eq}^{ij}(T) = \frac{B_j(T)-B_i(T)}{A_j(T)-A_i(T)}.
\end{equation}
The phase-field profile obtained with the remaining term 
of Eq. (\ref{mechanical}) and the given boundary 
conditions reads
\be
\label{tanh}
p_i(x)= \frac{1}{2}\left[1-\tanh\left(\frac{x-x_0}{\sqrt{2} W}\right)\right],
\ee
the desired solution for an interface centered on $x_0$. Equations
(\ref{mueq}) and (\ref{tanh}) together with the definition of $\mu$
[Eq. (\ref{muexplicit})]
and the condition $\mu=\mu_{\rm eq}^{ij}$ determine
the concentration profile,
\be
\label{cprof}
c(x)= \mu_{\rm eq}^{ij} + A_i g_i[\vec p(x)] + A_j g_j[\vec p(x)].
\ee
Taking the limits $x\to\pm\infty$, we find the concentration of
phase $i$ in equilibrium with phase $j$, 
\begin{equation}
\label{coexisting} 
c_i^{ij}=A_i+\frac{B_j-B_i}{A_j-A_i}.
\end{equation}
By choosing appropriate functions $A_i(T)$ and $B_i(T)$, we
can reproduce any phase diagram as characterized by its
coexistence lines $c_i^{ij}(T)$. Since
only free energy or concentration differences between
phases are relevant [note the form of the square bracket
in Eqs. (\ref{ponij}) or (\ref{mechanical})], 
we may choose $A_{\rm L}=B_{\rm L}=0$ 
without loss of generality. With the remaining four functions
of $T$ we can fit two liquidus and two solidus lines; 
the solid--solid equilibrium is then constrained, 
but this is irrelevant for the one-sided case, 
for which the solid has no dynamics.

\subsubsection{Relation to thermodynamics}
\begin{figure}
\centerline{\psfig{file=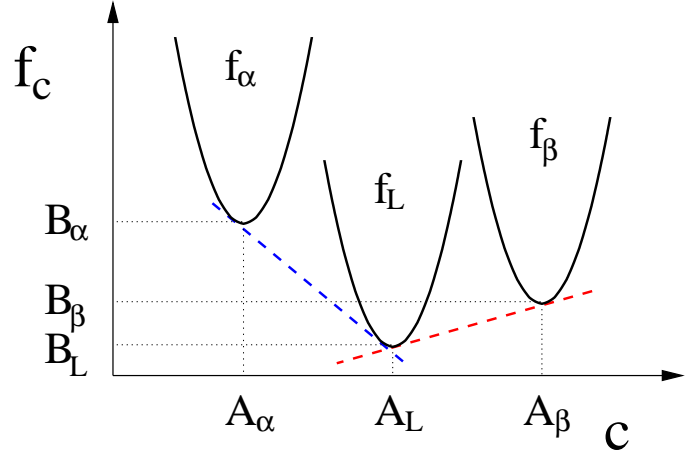,width=.5\textwidth}}
\vskip 1cm
\caption{The $f_c$ part of the free-energy density is an interpolation
between three parabolic bulk phase free energies; the equilibrium
compositions and chemical potentials can be obtained by the common 
tangent construction illustrated by the straight, dashed lines
for both solid--liquid equilibria.
This construction corresponds to the conditions of equal
chemical and grand potential in the two coexisting phases.}
\label{commontangent}
\end{figure}

To obtain the thermodynamic interpretation of the above
solvability condition $V(0)=V(1)$, note that
the two terms in the middle of Eq. (\ref{mechanical})
correspond to 
$(1/2) (d/dp_i) f_p (p_i,1-p_i,0)$
and $(1/2) \tilde{\lambda}(\partial/\partial p_i) f_c(p_i,1-p_i,0)$,
respectively \cite{footonehalf}. 
The second is a partial derivative because
$f_c$ also depends on $p_i$ implicitly through $c(p_i)$
[Eq. (\ref{cprof})]; its total derivative is
$(d/d p_i)f_c=(\partial/\partial p_i) f_c 
+ (\partial f_c/\partial c)(d c/ dp_i)$,
where $\partial f_c/\partial c = \mu$ is the constant 
$\mu_{\rm eq}^{ij}$ in equilibrium. Therefore,
this second term can also be written as
$(1/2)\tilde{\lambda}(d/dp_i)(f_c-\mu c)$.
By formal integration of the second equality in 
Eq. (\ref{mechanical}) we thus obtain (for constant $\mu$)
\be
V(p_i) = - \frac{1}{2} \left\{ f_p(p_i,1-p_i,0) 
+ \tilde{\lambda} \left [f_c(p_i,1-p_i,0,c) - \mu_{\rm eq}^{ij} c 
\right ] \right\}
\ee
up to a constant, to be compared to
the (dimensional) grand potential density,
\be
\omega = f -X \mu c = K f_{\rm grad} + H f_p + X (f_c - \mu c)
= K f_{\rm grad} -2H V(p_i),
\ee
where the last equality holds for constant $\mu$
and we recall that $\tilde{\lambda}=X/H$. 
Since $f_{\rm grad}$ vanishes in bulk phases,
the condition $V(0)=V(1)$ implies that the grand potential
density be equal in the two coexisting phases,
$\omega(0)=\omega(1)=\omega_{\rm eq}^{ij}$.
We hence see that the 
two conditions for the existence of a stationary solution
correspond to equality of chemical and grand potentials.
Since $f_p$ takes the same value in all bulk phases,
we note that the equality of grand potentials simply
requires that of $f_c-\mu c$. The latter, together with
a constant $\mu=\partial f_c/\partial c$, is what is usually
represented graphically
in the well-known common tangent construction, illustrated in 
Fig.~\ref{commontangent}.
%

\subsubsection{Surface tension and choice of $f_c$}
\label{choicefc}

At this point, it is important to realize that the conditions
of stationarity, or, equivalently, the common tangent construction,
only require the grand potential $\omega$ and the quantity
$f_c-\mu c$ to take equal values {\em at both sides} of the interface.
However, they both generally vary {\em through} it, which means that
their surface excesses are typically finite.
The surface excess of the grand potential, for instance,
gives the surface tension,
\be
\label{surftensdef}
\sigma_{ij}=\int_{-\infty}^{+\infty}
\{\omega[\vec p(x),c(x)]-\omega_{\rm eq}^{ij}\} dx.
\ee

The particularity of our model can now be understood:
At equilibrium, $\mu=\mu_{\rm eq}^{ij}$ happens to make
the square brackets and hence the whole term
proportional to $\tilde{\lambda}$ in Eq. (\ref{mechanical})
vanish for any value of $p_i$, i.e., 
throughout the whole interface. This decouples 
the phase and chemical potential fields at equilibrium,
as emphasized in the introduction. More precisely,
the equilibrium phase-field profile
[Eq.(\ref{tanh})] balances the derivatives of
$Kf_{\rm grad}$ and $Hf_p$ only; moreover,
since we saw that the term proportional to $\tilde{\lambda}$
corresponds to $(1/2)\tilde{\lambda}(\partial/\partial p_i)
f_c(p_i,1-p_i,0)=(1/2)\tilde{\lambda}(d/dp_i)(f_c-\mu c)$, the
fact that it vanishes means that
$f_c-\mu c$ is constant throughout the interface,
i.e., it has no surface excess. Therefore, it does not
contribute to the surface tension, which again is
determined purely by the remaining $Kf_{\rm grad}+Hf_p$. 
We obtain
\be
\label{sigmamin}
\sigma_{ij} = \frac{\sqrt{2}}{3} \sqrt{KH} 
= \frac{\sqrt{2}}{3} \; WH.
\ee

To better realize the importance of this point, it is
useful to consider the roles of the coefficients $K$,
$H$ and $X$ in the free energy. $X$ is the dimensional
prefactor of the concentration-dependent terms and
hence sets the magnitude of the thermodynamic driving
forces, which is a macroscopic, physically measurable
quantity. In contrast, $K$ and $H$ can be adjusted
in order to achieve a desired surface tension and interface
thickness. In our model, this is particularly straightforward
since $\sigma_{ij}$ does not depend on $X$: The
interface thickness $W$ can be varied for computational convenience
while keeping the measured value for $\sigma_{ij}$ just by fitting 
$H$ [see Eq. (\ref{sigmamin})], and, still, $X$ can be chosen 
independently to match its measured value. The capillary lengths 
$d_i$ are proportional to the ratio of the surface tension to the
driving forces ($d_i\propto\sigma_{i{\rm L}}/X$). Since 
$\sigma_{ij}\propto WH$, $d_i\propto WH/X=W/\tilde{\lambda}$, 
a scaling that will be confirmed by the thin-interface analysis 
given in the next subsection. This means that scaling up
simultaneously $W$ and $\tilde\lambda$ will leave the physics 
invariant, as desired.

For a generic model, in contrast, the terms proportional to $\tilde{\lambda}$ 
do {\em not} vanish in the equation of motion for $p_i$ nor
in the expression for the grand potential. Therefore,
the phase and chemical potential fields do not decouple; both
the phase-field profile (and therefore its thickness $W$) and the
surface tension depend on the three constants $K$,
$H$, and $X$, and no general analytic solution exists.
In the expression for the surface tension, there is an extra 
contribution coming from $f_c$, which, on dimensional grounds,
should be proportional to $WX$. Therefore, changing $W$ while
keeping the surface tension and driving forces fixed is far
more complicated. Furthermore, the coupling to $f_c$ will introduce 
some dependence of the surface tension and capillary lengths on the 
composition and temperature. While such dependencies may reflect
some physical effects, it is difficult to control them properly 
since they will be blown up as the interface thickness $W$ is 
scaled up to allow for simulations.

The advantage of our model comes from the fact
that the partial derivative of $f_c$ with respect to
any of the two phase fields present on an interface, 
$(\partial f_c/\partial p_i)|_{p_\alpha+p_\beta+p_{\rm L}=1, p_k=0}$,
vanishes at equilibrium {\em for any value of $p_i$}, i.e.,
thanks to the particular $f_c$ used. More specifically, 
we achieve this through an $f_c$ whose partial
derivative splits in (i) a factor which depends on $p_i$ only and 
(ii) one which depends on $\mu$ only, but none on $c$.
In turn, for the particular type of polynomial construction of
$f_c$ we use, (i) can be traced back to the symmetry of the $g_i$'s
[Eq.~(\ref{gsymmetry})],
and (ii) to the fact that the second derivatives of the
bulk-phase free energies with respect to concentration 
are constant and all equal.
The symmetry of the $g_i$'s can be understood physically as ensuring
that $f_c$ depends only on relative differences
of free energy and concentration between any pair of phases $i$ and $j$, 
{\em for any phase fraction} of the two phases $p_i$ and $1-p_i$.
This is reflected in the expression in square brackets in
Eqs. (\ref{ponij}) or (\ref{mechanical}).
Apart from this symmetry, the $g_i$ are arbitrary, phenomenological
phase-field interpolation functions. 

In contrast, the free
energies of the different bulk phases are, in principle, 
functions that are fixed for a given alloy system and
cannot be freely adjusted. Our free energy is therefore
an approximation, justified by the fact that it yields
the desired phase diagram. 
However, apart from the phase diagram, the free energy also
determines the latent heats, as we shall outline below,
and these enter the capillary lengths [Eq. (\ref{capillarydef})];
moreover, due to the Gibbs-Thomson effect or to kinetics,
the concentrations at both sides of a solid--liquid interface
can deviate from the prediction of the phase diagram at a
given temperature, which affects the amount of impurity
actually rejected.
As we shall see, our
approximation for $f_c$ constraints the capillary lengths and
slightly modifies the impurity redistribution; however, the
first is compatible with the experimentally measured values,
and the second is completely
negligible for low-speed solidification. To get rid of 
this approximation, it should be possible to use an approach 
recently introduced for dilute alloys in which
internal energy and entropy are interpolated by two different 
functions \cite{onesided}. However, the generalization of this 
method to two-phase solidification, that is, the introduction of 
two different sets of $g_i$ functions, is outside the scope of 
the present paper.

\subsection{Mapping of each solid--liquid interface 
to single-phase solidification}
\label{mapping}
Here, we address the behavior of our model for small but finite
values of the interface thickness $W$. 
It would be very interesting to investigate
this behavior including the triple junctions; however, the problem 
there is quite involved, since two independent phase fields and 
the chemical potential vary rapidly and are all coupled. Therefore, 
we limit ourselves to an analysis of the interfaces, taking 
advantage of a mapping to single-phase solidification. The behavior
of the trijunction points will be investigated numerically and 
discussed in detail in Sec.~\ref{numerics}.

We are actually interested in solid--liquid interfaces,
since the solid--solid one has no dynamics. We hence set
$j=\rm L$ in Eqs. (\ref{ponij}) and (\ref{muonij}).
To proceed, we
rewrite the expression
$\mu(A_{\rm L}-A_i) - (B_{\rm L}-B_i)$ 
thus appearing in Eq. (\ref{ponij}) as 
$(A_{\rm L}-A_i)(\mu - \mu_{\rm eq}^{i{\rm L}})$, where 
$\mu_{\rm eq}^{i{\rm L}}=(B_{\rm L}-B_i)/(A_{\rm L}-A_i)$
is the dimensionless chemical potential for coexistence of the
solid $i$ and the liquid as
given by Eq. (\ref{mueq})].
It hence becomes apparent that the driving force,
$-15\tilde\lambda p_i^2(1-p_i)^2 (A_{\rm L}-A_i)(\mu - \mu_{\rm eq}^{i{\rm L}})$, 
is proportional to a deviation from equilibrium.
Let us first consider isothermal solidification, so that
$A_i(T)$, $A_{\rm L}(T)$ and $\mu_{\rm eq}^{i{\rm L}}(T)$
are just constants.
Then, the evolution equations can we rewritten in terms
of a new variable
\begin{equation}
\label{change}
u = (\mu - \mu_{\rm eq}^{i{\rm L}})/(A_{\rm L}-A_i),
\end{equation}
which measures departure from equilibrium on the $i$--L interface,
and a new phase field $\phi_i \equiv 2p_i -1$, 
which takes the values $+1$ and $-1$ in the $i$ solid and liquid phases,
respectively. We obtain:
\begin{eqnarray}
\label{phionil}
\tau(\vec p)\frac{\partial\phi_i}{\partial t} =
  W^2 \nabla^2\phi_i
+ \phi_i (1-\phi_i^2 ) - \frac{15}{8} (A_{\rm L}-A_i)^2 \tilde\lambda (1-\phi_i^2)^2 u \\
\label{uonil}
\frac{\partial u}{\partial t} =
\vec\nabla\cdot \left[D(\vec p)
				   \vec\nabla u\right]
+\frac{\partial g_i(p_j=0)}{\partial t},
\end{eqnarray}
where $j$ denotes the other (absent) solid phase.

With a constant relaxation time, $\tau(\vec p) \equiv \tau_0$, and a 
constant diffusivity, $D(\vec p) \equiv D$, the above equations constitute
precisely the phase-field model for the solidification of a pure substance
treated by Karma and Rappel, if the combination 
$(15/8) (A_{\rm L}-A_i)^2 \tilde\lambda$
is identified with the coupling constant $\lambda$ used in Ref.~\cite{kr};
more precisely, we recover the variational version of their model.
Therefore, we can use their results on its thin-interface behavior,
i.e., the behavior for values
of $W$ much smaller than a typical lengthscale of the microstructural
pattern. This is described by an effective free
boundary problem for the field $u$. Undoing the change of variables of
Eq. (\ref{change}) and applying Eq. (\ref{muexplicit}), 
we can compare it with the desired free-boundary
problem of Eqs. (\ref{fbpeqs}):

We find that the Gibbs-Thomson boundary
condition, Eq. (\ref{githo}), is satisfied, with capillary lengths
and kinetic coefficients related to the phase-field parameters by
\begin{eqnarray}
\label{capillary}
d_i & = & a_1 \frac{W}{|A_{\rm L}-A_i|\tilde\lambda}, \\
\label{kinetic1}
\beta_i & = & a_1 \left [ \frac{\tau_0}{|A_{\rm L}-A_i|\tilde\lambda W}
				- a_2|A_{\rm L}-A_i|\frac{W}{D} \right ],
\end{eqnarray}
where $a_1=\sqrt{2}/3$ and $a_2=0.7464$ are numerical constants 
related, but not equal to \cite{foot2} those in Ref.~\cite{kr}.
We note that, in general, the expressions for the capillary lengths
and the kinetic coefficients are not the same for the two solid--liquid
interfaces, but depend on the $A_i$'s, 
and hence on concentration differences
[recall Eq. (\ref{fc})].
As anticipated, the parameter $\tilde\lambda$ controls the ratio of the interface 
thickness $W$ to the smallest physical lengthscales, the capillary lengths $d_i$. 
The former can be varied while keeping the latter fixed just by changing
$\tilde\lambda$ accordingly. For quantitative output, the simulation
results should hence become independent of $\tilde\lambda$ for 
$\tilde\lambda$ small enough. This will be checked for our model 
in Sec.~\ref{numerics}. Note that, as discussed in detail in
Refs.~\cite{kr,onesided}, convergence may be achieved for $\tilde\lambda$ 
much larger than $1$, since it is only assumed that $W$ is much smaller than 
a typical length scale of the pattern, which, in turn, is usually much 
larger than $d_i$.

The diffusion equation, Eq. (\ref{diffusion}), is also satisfied.
As for the mass conservation condition, we find
\be
\label{stefansymmetric}
D \hat n \cdot (\vec\nabla c|_+-\vec\nabla c|_-) = (A_i-A_{\rm L}) v_n,
\ee
where $\vec\nabla c|_+$ and $\vec\nabla c|_-$ denote the concentration
gradients on the liquid ($+$) and the solid ($-$) side of the 
interface, respectively. There are two differences with the
Stefan condition of the desired FBP, Eq.~(\ref{mc}). Obviously, in the
one-sided case the diffusion flux on the solid side vanishes,
and hence the second term on the left hand side is absent.
We address this case in Sec.~\ref{1sided}.
But there is a second difference: The right hand side of
Eq.~(\ref{stefansymmetric}) corresponds to the right hand side 
of Eq.~(\ref{mc}) only for constant concentration gaps, $k_i=1$.
To see this, substitute Eq.~(\ref{gibbsthomson}),
absorbing its temperature dependence in the phase-diagram concentration
gaps $c_i^{i{\rm L}}-c_{\rm L}^{i{\rm L}}$, into Eq.~(\ref{mc}).
This yields
$D\hat n\cdot\vec\nabla c = 
v_n \left [ c_i^{i{\rm L}}-c_{\rm L}^{i{\rm L}}
		  \pm (1-k_i)(d_i\kappa + \beta_i v_n) \right ]$.
The term $c_i^{i{\rm L}}-c_{\rm L}^{i{\rm L}}$ corresponds to our $A_i-A_{\rm L}$
[see Eq. (\ref{coexisting})], but the other terms are lacking.
This is due to the use of equal
$\partial^2 f_{c,i}/\partial c^2$ for all three bulk-phase 
free-energy functions $f_{c,i}$ [Eq. (\ref{fci})], 
as announced in Sec.~\ref{equilibrium}:
When the chemical potential
is shifted from its temperature-dependent equilibrium value
for a flat interface,
the corresponding shifts in concentration are determined
by $\partial\mu/\partial c=\partial^2 f_c/\partial c^2$ and
are hence the same at both sides of each solid--liquid interface.
This explains why our model displays the right deviations of $c$
from its phase-diagram value at the liquid side 
[correct Gibbs-Thomson condition, Eq. (\ref{githo})],
but does not reflect them in the concentration gaps that
appear in the impurity rejection.  
However, the magnitude of the missing terms is small, since the 
departure from the equilibrium phase diagram is very small in slow solidification:
The $\beta_i v_n$ term is usually neglected and will be
set to zero in our simulations; as for $d_i\kappa$, 
a typical curvature $\kappa$ is given by the inverse of the 
lengthscale of the microstructural pattern. In turn, this lengthscale
goes as $\sqrt{\bar d l_D}$,
where $l_D=D/V_p$ is the diffusion length.
Therefore, $d_i\kappa\sim \sqrt{\bar d/l_D}$. 
For typical experimental values of slow solidification
$\bar d/l_D \sim 10^{-4}-10^{-5}$, so this correction can be safely
neglected. 

The other constraint coming from the equal $\partial^2 f_{c,i}/\partial c^2$
is reflected in Eq.~(\ref{capillary}): The ratio of the capillary
lengths is fixed to
\be
\label{constraintminimal}
\frac{d_\alpha}{d_\beta} = \frac{|A_\beta-A_{\rm L}|}{|A_\alpha-A_{\rm L}|}.
\ee
This relation can be understood from thermodynamic considerations:
The latent heats of the two solid--liquid phase transformation
can be evaluated from the bulk-phase free energies $f_i$ 
of [Eq. (\ref{fci})] by 
$L_i=T(s_{\rm L}-s_i)$, where $s_i=-(\partial f_i/\partial T)|_c$ is the 
specific entropy of phase $i$ (and similarly for the liquid).
Making use of the conditions of equal chemical and grand 
potentials, we find
$L_i=(\partial\mu_{\rm eq}^{i\rm L}/\partial T)|_c 
(c_i^{i\rm L}-c_{\rm L}^{i\rm L})
= (\partial\mu_{\rm eq}^{i\rm L}/\partial c) 
|c_i^{i\rm L}-c_{\rm L}^{i\rm L}|/m_i$.
Since for both solid--liquid equilibria 
$(\partial\mu_{\rm eq}^{i{\rm L}}/\partial c)=
\partial^2 f_c/\partial c^2|_{\rm eq}$
are the same, we find
$L_\alpha/L_\beta = 
|(c_\alpha^{\alpha\rm L}-c_{\rm L}^{\alpha\rm L})/
(c_\beta^{\beta\rm L}-c_{\rm L}^{\beta\rm L})|
(m_\beta/m_\alpha)$. 
Using the definition of the capillary lengths, Eq. (\ref{capillarydef}), 
this yields
\begin{equation}
\label{thermoconstraint}
\frac{d_\alpha}{d_\beta} = 
\frac{\sigma_{\alpha\rm L}}{\sigma_{\beta\rm L}}
\left | \frac{c_\beta^{\beta\rm L}-c_{\rm L}^{\beta\rm L}}
	        {c_\alpha^{\alpha\rm L}-c_{\rm L}^{\alpha\rm L}}
\right | =
\frac{\sigma_{\alpha\rm L}}{\sigma_{\beta\rm L}}
\left | \frac{A_\beta-A_{\rm L}}{A_\alpha-A_{\rm L}}
\right |,
\end{equation}
where the second equality makes use of Eq. (\ref{coexisting}).
Since, in our minimal model, $\sigma_{\alpha {\rm L}}=\sigma_{\beta {\rm L}}$,
Eq.~(\ref{constraintminimal}) follows. In general, the ratio depends 
on the temperature through the $c_i^{ij}(T)$. Near the eutectic 
point where solidification occurs for small undercoolings,
\begin{equation}
\label{thermoconstraintatte}
\frac{d_\alpha}{d_\beta} \approx
\frac{\sigma_{\alpha {\rm L}}}{\sigma_{\beta {\rm L}}}
\left | \frac{c_\beta}{c_\alpha} \right |.
\end{equation}
This latter relation is satisfied by the reference alloy of our present study, 
CBr$_4$--C$_2$Cl$_6$, to within experimental accuracy.

Note that these lacking terms in the solute rejection and
the restriction on the ratio of the capillary lengths 
do not vary with the interface thickness, 
but are zeroth-order corrections:
They are not introduced by the diffuseness of the interfaces,
but by our approximation for the free energy, which can
be replaced by more sophisticated choices if needed, as 
already discussed at the end of Sec.~\ref{choicefc}.

Let us conclude this section with a comment on directional
solidification, where the assumption of constant $A_i(T)$ 
and $A_{\rm L}(T)$ is no longer valid. If we still want to 
apply the mapping to single-phase solidification, we must 
neglect the temperature variation of $A_i(T)$ and $A_{\rm L}(T)$
within the region where the mapping holds.
The scale of variation of the temperature is set by the thermal lengths
$l_T^i$ defined in Sec.~\ref{fbp}, whereas the mapping region needs to 
extend over at least a few times $W$ in order for the thin-interface 
behavior of the model to be well defined (and the results of Ref.~\cite{kr}
to apply). Because these results already assume that the lengthscale of
the pattern is much larger than $W$, and taking into account that the
thermal lengths are typically far larger than this pattern lengthscale,
the condition $W\ll l_T^i$ is automatically met. Therefore, even in
presence of a temperature gradient, the results of Eqs.~(\ref{capillary}) 
and (\ref{kinetic1}) apply, with the value of $T$ (and hence $A_i(T)$ and 
$A_{\rm L}(T)$) taken in the center of the interface.

\subsection{Independent kinetic coefficients}
\label{indkinetics}

The relation between the phase-field parameters $\tau_0$, $\tilde{\lambda}$
and $W$ and the interface
kinetic coefficients of the FBP is given by Eq.~(\ref{kinetic1}),
where
$|A_{\rm L}-A_i|=|c_{\rm L}^{i {\rm L}}-|c_i^{i\rm L}|$ are set by the 
phase diagram, $D$ is a material parameter, and $a_1$ and $a_2$ are
numerical constants. Given a desired accuracy as fixed by $W$
or, equivalently, $\tilde{\lambda}$ [linked by Eq.~(\ref{capillary})], 
the only free parameter left is the phase-field relaxation time 
$\tau_0$. However, we need at least two free parameters in the model, 
since there are two kinetic coefficients that are {\em a priori} 
independent. In the formulation of 
Ref.~(\cite{interface}) with a double-obstacle potential, this 
is implemented by specifying different relaxation times $\tau_{ij}$ 
for each binary ($i$--$j$) phase transformation. 
In our formulation with a smooth free 
energy functional, this is not possible, as shown in 
Appendix \ref{timeconstants}. Instead, we achieve independent kinetic
coefficients by keeping the same relaxation time $\tau$ for all possible
transformations, 
{\em but} making $\tau$ dependent on the phase fields.

In order to maintain the mapping to the model of Ref.~\cite{kr} on
the individual interfaces, $\tau$ has to be a constant along
those interfaces. A function taking different, constant values on all 
three interfaces necessarily has discontinuity lines inside 
the Gibbs simplex. Here, we are mainly interested in controlling
the dynamics of the solid--liquid interfaces, since, in the final
one-sided model, the solid--solid interface does not move. Therefore, 
we use a function that takes constant values 
($\tau_\alpha$ and $\tau_\beta$) on the two solid--liquid interfaces
only, and interpolates smoothly between them {\em inside} 
the Gibbs simplex:
\begin{equation}
\label{tauofp}
\tau(\vec p) = 
\left\{
\begin{array}{lll}
\bar\tau
+ \frac{\tau_\beta-\tau_\alpha}{2} 
\frac{(p_\beta-p_\alpha)}{(p_\alpha+p_\beta)}
& \mbox{if} & p_{\rm L}\neq 1 \\
\bar\tau & \mbox{if} & p_{\rm L}=1, \\
\end{array}
\right.
\end{equation}
where $\bar\tau=(\tau_\alpha+\tau_\beta)/2$.
Note that the only discontinuity is on the vertex
$p_{\rm L}=1$.
It did not cause any problems in practice, but, 
if needed, it could be smoothed 
out in a small neighborhood of the vertex without inducing appreciable
errors in the calculations.

As a result of this procedure, the two kinetic coefficients,
\be
\label{kinetic}
\beta_i = a_1 \left [ \frac{\tau_i}{|A_{\rm L}-A_i|\tilde\lambda W}
						- a_2|A_{\rm L}-A_i|\frac{W}{D} \right ],
\ee
can now be adjusted independently by choosing
the two constants $\tau_\alpha$ and $\tau_\beta$.

\subsection{Non-variational model}
\label{nonvariational}

It has been shown for phase-field models of single-phase
solidification \cite{kr,almgren} that it is often advantageous 
to introduce an additional degree of freedom by switching to a 
non-variational formulation, which exploits the two
different roles of the tilting functions $g_i$:
In the evolution equations for the phase fields,
Eq. (\ref{evolutionofpmin}), 
they favor one bulk phase over the other; 
in the one for the chemical potential $\mu$, 
Eq. (\ref{evolutionofmumin}), they constitute
a source or sink of $\mu$, which corresponds to
impurity rejection or adsorption, respectively.
The $g_i$ need to satisfy certain common requirements for both of
their roles (although for different reasons), 
namely to interpolate from 0 to 1 as $p_i$ goes from
0 to 1 and to be antisymmetric with respect to the point $p_i=p_j=1/2$ on
$i$--$j$ interfaces [Eq.~(\ref{gsymmetry0})]. However, the requirement 
of flatness, Eq. (\ref{flatness}), constrains only the derivatives
with respect to the phase fields, which do not enter the
evolution equation of the chemical potential $\mu$. 

Therefore, we can switch to a different chemical-potential-like
variable,
\be
\label{munewdef}
\mu = c- \sum_i A_i h_i(\vec p),
\ee
which amounts to replace the $g_i$ in Eq.~(\ref{muexplicit})
by new functions $h_i$ that have the same limits and symmetries
as the $g_i$, but do not necessarily satisfy Eq. (\ref{flatness}).
This implies that the $\partial_t g_i$ in Eq.~(\ref{evolutionofmumin})
are replaced by $\partial_t h_i$. Furthermore, 
a different expression for $f_c$ has to be used to derive the
phase-field equations, namely
\be
\label{fcnv}
f_c=\sum_i g_i(\vec p) [B_i(T)-\mu A_i(T)],
\ee
which is the form given in our preliminary account \cite{Folch03}.
The mapping to single-phase solidification can be repeated:
For the choice $h_i=p_i$, the result corresponds exactly to
the isothermal variational model of Ref.~\cite{kr}; as a result,
Eqs.~(\ref{capillary}) and (\ref{kinetic}) hold again, with the
same value of $a_1$ as before, but now with $a_2=1.175$.

The choice $h_i(\vec p)=p_i$ is quite advantageous for simulations,
because the source term in Eq.~(\ref{evolutionofmumin}) is less 
peaked inside the interface than for the fifth-order 
polynomial $g_i(\vec p)$, which makes it possible to use a 
coarser discretization, resulting in a considerable gain in 
computational speed, as discussed in detail in Ref.~\cite{kr}.

\subsection{One-sided model}
\label{1sided}

We now address the one-sided case, for which the solute diffusivity 
in the solids is neglected compared to that of the liquid $D$.
This is much more realistic for alloy solidification than the 
symmetric model considered up to now. We hence set
\begin{equation}
\label{mobility}
D(\vec p) = D q(p_{\rm L}),
\end{equation}
with a function $q(p_{\rm L})$ that interpolates between $1$ in the
liquid and $0$ in the solid. 

This seemingly small change has important consequences. We already
mentioned in Sec.~\ref{fbp} that a strictly vanishing solid 
diffusivity prevents solid--solid interfaces to move. For the
solid--liquid interfaces, the introduction of a phase-dependent
diffusivity breaks the symmetry of the impurity diffusion between 
solid and liquid. Furthermore, the variation of the diffusivity 
$D(\vec p)$ through the diffuse solid--liquid interfaces needs
to be considered in the asymptotic analysis, which becomes
quite involved \cite{onesided,almgren}; we will only resume
the most important points here.

It turns out that, in general, the effective 
free-boundary problem (FBP) obtained 
from the thin-interface analysis displays several terms that
are not present in the original one
and that scale with the interface thickness $W$. As explained 
in detail in Ref.~\cite{onesided}, in analogy with the thermodynamics 
of diffuse interfaces these terms can be linked to 
``surface excesses'', which are the integral through the interface
of the excess of a quantity, where excess means the difference 
of its actual value at a point inside the diffuse interface region 
and its reference bulk value at the nearest side of the interface 
(the bulk values in the solid and the liquid might differ). 

Let us consider the evolution equation for the chemical potential on
the solid--liquid interfaces, Eq. (\ref{muonij}), where the $g_i$ were
replaced by $h_i$ as explained in the previous subsection. Three relevant
surface excesses have been identified. The first is that of the ``source
function'' $h_i(\vec p)$: since it determines the equilibrium solute
profile, an excess in this quantity confers a net impurity content 
to the interface (solute adsorption at the interface). The quantity
of adsorbed solute increases with the area of the interface, which 
modifies the mass conservation condition, Eq. (\ref{mc}), at moving 
curved interfaces (interface stretching). For the 
one-sided model, we have two more surface excesses:
that of diffusivity $D(\vec p)$, 
which leads to solute diffusion along the interface (surface diffusion)
and thus also modifies the mass conservation condition; and that 
of the ratio $h_i(\vec p)/D(\vec p)$, which
generates the well-known solute-trapping effect and
shows up in the modified FBP in the form of a
jump of the chemical potential through the interface \cite{onesided}. 

For the symmetric model in which $D(\vec p)$ is actually a constant, 
the choice of $h_i(\vec p)$ odd with respect to the center of the 
interface ensures both the absence of interface stretching and of 
solute trapping. However, for the one-sided case, one has to 
eliminate surface diffusion too. The easiest way to do this is to also
choose $D(\vec p)$ odd. The simplest possible interpolation,
$q(p_{\rm L})=p_{\rm L}$, already satisfies this condition, and
will be adopted in the following. The problem, however, is then 
that the surface excess of $h_i(\vec p)/D(\vec p)$ does not vanish, 
so that solute trapping is present.

With only two adjustable interpolation functions, one can
generally not eliminate all three undesired surface excesses; 
more freedom is needed in the formulation of the model.
One way to obtain this is to add a term counterbalancing one of 
the three effects, while choosing $h_i(\vec p)$ and $D(\vec p)$
to eliminate the other two. Here, we generalize to two-phase 
solidification one possible such term, originally introduced 
for single-phase solidification in Ref.~\cite{onesided}:
We add to the solute current a new phenomenological 
contribution, an ``antitrapping current'' which drives the 
otherwise trapped solute into the liquid phase. 

By analogy with Ref.~\cite{onesided}, 
its form on each solid--liquid interface is easy
to determine: It is directed parallel to the outward normal
of the solid--liquid interface, and its magnitude is
proportional to the interface thickness $W$ and the 
intensity of the solute release which caused the trapping, 
$(A_{\rm L}-A_i)\partial h_i/\partial t$.
How to interpolate this current between the two solid--liquid 
interfaces is less clear, since no asymptotic analysis or mapping 
to single-phase solidification is available near a trijunction.
We numerically tested various possibilities for interpolations,
and chose the one that yielded the best convergence properties.
It can be heuristically motivated as follows: The outward normal
to each phase is given by $\hat n_i=-\vec\nabla p_i/|\vec\nabla p_i|$.
Let us consider $\hat n_\alpha$. On the $\alpha$--liquid interface,
it is antiparallel to $\hat n_{\rm L}$, since $p_\alpha=1-p_{\rm L}$;
in contrast, on the $\alpha$--$\beta$ interface it is antiparallel
to $\hat n_\beta$, since $p_\alpha=1-p_\beta$. Upon growth,
the $\alpha$ front advances along the direction of $\hat n_\alpha$
with a speed proportional to $\partial p_\alpha/\partial t$.
The antitrapping current should always be directed towards the 
liquid, even if the solid formed is a ``mixture'' of both solid 
phases as happens in a trijunction; therefore, we choose 
to direct it along $\hat n_{\rm L}$ throughout the whole system.
However, only the component of the growth speed directed
toward the liquid should contribute to the magnitude of the
antitrapping current; therefore, we multiply the source strength
for each phase by the scalar product $-\hat n_{\rm L}\cdot \hat n_i$, 
which is equal to $1$ on the solid--liquid interfaces, but smaller 
than $1$ inside the trijunction.

The total concentration current thus reads
\begin{equation}
\label{totalcurrent}
\vec J = -Dp_{\rm L}\vec\nabla\mu 
+ 2aW \hat n_{\rm L}
\sum_{i=\alpha,\beta} (A_{\rm L}-A_i) (-\hat n_{\rm L}\cdot \hat n_i)
      \frac{\partial p_i}{\partial t},
\end{equation}
where $2a$ is a prefactor to be adjusted, and we have taken $h_i=p_i$. 
Indeed, the particular form of the anti-trapping current given here is 
to be used {\em only} in conjunction with the simplest choices both for the 
diffusivity, as given in Eq. (\ref{mobility}) with $q(p_{\rm L})=p_{\rm L}$, 
and the source function, $h_i=p_i$; see Ref. \cite{onesided} for further details. 
 
With this new concentration current [Eq. (\ref{totalcurrent})],
we rederive Eq. (\ref{uonil}):
\begin{eqnarray}
\label{newuonil}
\frac{\partial u}{\partial t} & = &
D\vec\nabla\cdot \left ( \frac{1-\phi_i}{2}\vec\nabla u\right )
+ \frac{1}{2} \frac{\partial \phi_i}{\partial t} 
+ aW \vec\nabla\cdot
\left (\frac{\partial \phi_i}{\partial t}
\frac{\vec\nabla\phi_i}{\left |\vec\nabla\phi_i\right |}\right ).
\end{eqnarray}
We recover the quantitative phase-field model for one-sided 
solidification with a constant concentration gap given in 
Ref.~\cite{onesided}. Again, this has the advantage that we do 
not need to analyze the small-$W$ behavior. From Ref.~\cite{onesided} we learn 
that the results of Eqs. (\ref{capillary}) and (\ref{kinetic}) for $d_i$ and
$\beta_i$ and the discussion thereafter still apply, and that 
$a=1/(2\sqrt{2})$ 
is the value for which the term exactly counterbalances solute trapping. 
Interestingly, the values of $a_1$ and $a_2$ stay also 
the same as given before, for reasons explained in Ref.~\cite{onesided}. 
The fact that the only quantitative model for one-sided
solidification available so far uses $h_i=p_i$ and that this
choice enables one to recover the same numerical values for $a_1$ and $a_2$ 
as in the case of constant diffusivity is obviously another reason, 
on top of its lower numerical cost, to use $h_i=p_i$ instead of $h_i=g_i$.

\subsection{Unequal surface tensions}
\label{unequal}

As we have seen in Sec. \ref{choicefc}, thanks to the particular form of 
our coupling $f_c$ between phase and concentration fields, the 
surface tension, or equilibrium surface excess of the grand
potential $f-\mu c$, reduces to the surface excess 
of the remaining two terms: $f_{\rm grad}$ and $f_p$.  
Since so far both $f_{\rm grad}$ 
and $f_p=f_{\rm TW}$ are symmetric with respect to the exchange 
of any two phases, the surface tensions of all 
interfaces are equal. In order to treat unequal surface tensions, 
this symmetry needs to be broken by modifying either of these two terms.
In multi-phase-field models, different gradient energies are 
used for each interface~\cite{mpfwith2obstacle}. In our approach, we want to 
maintain at least the condition of flatness, Eq.~(\ref{flatness}),
and it turns out to be easier to modify the potential part
(see Appendix \ref{generalgradients} for a detailed discussion).

We add a new term $f_{\rm saddle}$ to this potential part,
\be
\label{fpcomplete}
f_p = f_{\rm TW} + f_{\rm saddle},
\ee
\begin{equation}
\label{wholesaddle}
f_{\rm saddle}=\sum_i f_{{\rm saddle},i}.
\end{equation}
$f_{\rm saddle}$ is due to change the height of the saddle points on each binary 
interface by a tunable, different amount, through
the elementary functions $f_{{\rm saddle},i}$, which, in turn, 
should shift the saddle of 
$f_{\rm TW}$ separating phases $j$ and $k$ $(i\neq j,k)$. 
Therefore, $f_{{\rm saddle},i}$ should vanish on interfaces other than 
the $j$--$k$ one, and respect the valley character of all interfaces.

\begin{figure}
\centerline{\psfig{file=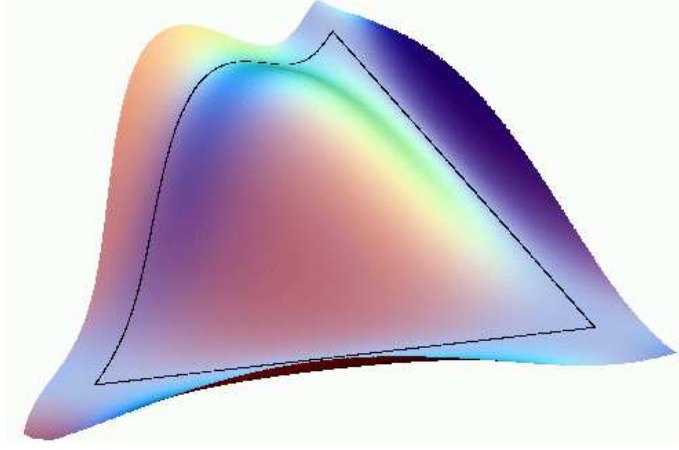,width=0.5\textwidth}}
\caption{Function used to lift the saddle between two phases 
and hence alter the surface tension. It is flat along the
two unaltered interfaces. The position and strength of the
hill in the middle are controlled by the parameter $b$ in
Eq.~(\protect\ref{saddle}) (here, $b=12$).
}
\label{saddleplot}
\end{figure}

A function positive everywhere that vanishes on all sides
automatically satisfies the latter two requirements, but obviously 
does not shift any saddle. However, it will be useful later on. 
The simplest such function is
\begin{equation}
\label{hill}
f_{\rm obs}=p_1^2p_2^2p_3^2, 
\end{equation}
which corresponds to an elevation (obstacle) on the triple point
and outside the Gibbs simplex. 
By dropping the $p_i^2$ factor we get a function which actually 
raises the saddle on the $j$--$k$ interface and still vanishes 
and has valleys on the others, but which is not flat on the $j$--$k$ 
interface in the direction perpendicular to it. We hence make the
ansatz $f_{{\rm saddle},i}=p_j^2p_k^2\tilde{f}_{{\rm saddle},i}(\vec p)$
and impose flatness at $p_i=0$. The resulting condition for 
$\tilde f_{{\rm saddle},i}$  reads 
$3(\partial \tilde{f}_{\rm saddle}/\partial p_i)|_{p_\alpha+p_\beta+p_{\rm
L}=1} 
=2\tilde{f}_{\rm saddle}/(p_jp_k)$
at $p_i=0$, so that the simplest choice turns out to be
$\tilde{f}_{{\rm saddle},i} = 2p_jp_k + 3p_i$. This actually 
corresponds to a function that is small everywhere but in the 
neighborhood of the $j$--$k$ saddle, where it is maximal. 
Therefore, its $j$--$k$ side is flat but concave.
To correct this concavity, we add to it the obstacle 
function $f_{\rm obs}$ given above,
\begin{equation}
\label{saddle}
f_{{\rm saddle},i}=a_i p_j^2p_k^2( 2p_jp_k + 3p_i + b p_i^2),
\end{equation}
through the term in $b$.
This composite function now has a valley that also runs 
along the $j$--$k$ side as long as $b > 9/2$. The larger $b$,
the closer the maximum of this function is to the triple point and 
the further to the $j$--$k$ saddle. In Fig. \ref{saddleplot}, 
we show this function for $b=12$. 
%
%

On any purely binary interface $p_i=0$, the whole function 
$f_{\rm saddle}$ reduces to $2a_i p^3(1-p)^3$, 
where $p$ is any of the other phase fields.
If we now put together the triple-well potential and the saddle
functions, we find a free-energy density
$f_p=f_{\rm TW} + f_{\rm saddle} = 2p^2(1-p)^2  [1 + a_i p(1-p)]$ 
on such interfaces. 

It is important to note that this
modification does not affect the coupling between phase fields
and concentration, and hence all the equilibrium compositions
and the chemical potential remain the same as in the minimal
model. In contrast, the surface tension is modified. Making use of 
the equipartition relation (i.e., the fact that $f_{\rm grad}=f_{\rm TW}$ 
for an equilibrium interface solution), we find the total surface excess
per unit area of $f_{\rm grad} + f_{\rm TW}$ to be
\begin{equation}
\label{tensions}
\sigma_{jk} = 2\sqrt{2}WH \int_{0}^{1} p(1-p)\sqrt{1 + a_i p(1-p)} dp, 
\;\; i\neq j,k.
\end{equation}
Of course, for $a_i=0$ the result reduces to Eq.~(\ref{sigmamin}).
For $a_i\neq 0$, it is straightforward to evaluate this integral
numerically and to tune the different surface tensions by using 
different $a_i$. However, the equilibrium profile of a $j$--$k$
interface whose saddle has been shifted by a finite $a_i$
is not any more the usual hyperbolic tangent solution of 
Eq.~(\ref{tanh}), but a kink-shaped profile that has to be
calculated numerically. Since the equilibrium profile is the
starting point for the entire asymptotic analysis, all calculations,
and in particular the determination of the numerical constants
$a_1$ and $a_2$ that appear in Eqs.~(\ref{capillary}) and (\ref{kinetic}),
have to be repeated for this new profile, taking also into
account the extra terms that are generated by $f_{\rm saddle}$
in the equations of motion. Similarly, the form of the antitrapping 
current needs to be adapted, using the methods of Ref.~\cite{onesided}. 
While this procedure is, in principle, straightforward, it is outside 
the scope of the present paper.

At this point, it is important to realize that the two solid--liquid
surface tensions in many eutectic alloy systems are quite similar,
and only the solid--solid surface tension is markedly different.
This is, for instance, the case in \alloy. Therefore, using equal 
$\alpha$--liquid and $\beta$--liquid surface tensions is usually
a good approximation; only the solid--solid surface tension 
needs to be modified, and hence we have $a_\alpha=a_\beta=0$,
but $a_{\rm L}\neq 0$. Then, we recover the quantitative
phase-field model on both solid--liquid interfaces; the solid--solid 
interface is not a problem, because 
it does not move (except close to the trijunction) in the one-sided model.

We should mention here that the computational effort to simulate the 
situation with unequal surface tensions ($a_k> 0$ for any $k$) is 
generally larger than for equal surface tensions ($a_k=0 \; \forall k$). 
This is due to the fact that the free-energy landscape of a $i$--$j$ interface 
whose saddle has been risen ($a_k>0,\;\; k\neq i,j$) is steeper
than that of an unaltered interface, so that a smaller grid spacing 
$\Delta x$ is necessary to discretize it in a convergent manner.

\section{Numerical tests}
\label{numerics}

\subsection{Implementation}
Our goal here is to show how the model derived in the preceding 
section is used in practice and to assess its validity and precision. 
We start by writing down the complete evolution equations for the 
one-sided, non-variational model with antitrapping current.
They are essentially the same as Eqs.~(\ref{evolutionofpmin})
and (\ref{evolutionofmumin}) of the minimal model, except that
Eq. (\ref{evolutionofpmin}) now includes a contribution
from $f_{\rm saddle}$, $D(\vec p)=Dp_{\rm L}$, the $g_i$ have 
been replaced by $h_i=p_i$ in Eq. (\ref{evolutionofmumin}) but 
not in Eq. (\ref{evolutionofpmin}),
and the total concentration 
current is now given by Eq.~(\ref{totalcurrent}),
which includes the antitrapping term. We make the equations
dimensionless by scaling lengths by $W$ ($\tilde x=x/W$, $\tilde z=z/W$,
$\tilde{\vec\nabla}=W\vec\nabla$) 
and times by the $\bar\tau=(\tau_\alpha+\tau_\beta)/2$
used in the equation that sets $\tau(\vec{p})$,
Eq. (\ref{tauofp}) ($\tilde t=t/\bar\tau$). 
The result reads
\begin{eqnarray}
\tilde\tau(\vec p)\frac{\partial p_i}{\partial \tilde t} & = &
   \tilde{\nabla}^2p_i
+ \frac{2}{3}\left [ -2p_i(1-p_i)(1-2p_i) + \sum_{j\neq i} p_j(1-p_j)(1-2p_j) \right ] 
  \nonumber \\
& & \mbox{}
+ 2      \left       \{ a_i p_jp_k \left [ (p_j+ p_k) f_{a,i} - 2 f_{b,i} \right ]
+ \sum_{j\neq i} a_j p_ip_k \left [ (p_i-2p_k) f_{a,j} +   f_{b,j} \right ]   \right \} 
  \nonumber \\
& & \mbox{}
+ \tilde\lambda\sum_j 
\left.\frac{\partial g_j}{\partial p_i}\right |_{p_\alpha+p_\beta+p_{\rm L}=1}
\left ( \mu A_j - B_j \right ) 
    \quad \forall i, 
\label{evolutionofp} \\
\frac{\partial \mu}{\partial \tilde t} & = &
\tilde{\vec\nabla}\cdot \left [\tilde D p_{\rm L}\tilde{\vec\nabla}\mu\right ]
   -\sum_i A_i \frac{\partial p_i}{\partial \tilde t}  \nonumber \\
 & & \mbox{} + 2a\sum_{i=\alpha,\beta} (A_i-A_{\rm L})(-\hat n_{\rm L}\cdot\hat n_i)\tilde{\vec\nabla}\cdot 
\left (\hat n_i\frac{\partial p_i}{\partial \tilde t}
\right ),
\label{evolutionofmu}
\end{eqnarray}
where $f_{a,k} \equiv p_ip_j + p_k (1+bp_k/3)$,
$f_{b,k} \equiv p_ip_j (1/2+bp_k/3)$, and we recall that
the $(\partial g_j/\partial p_i)|_{p_\alpha+p_\beta+p_{\rm L}=1}$
are given by Eqs. (\ref{gipi}) and (\ref{gjpi}),  
that $\hat n_i=\vec\nabla p_i/|\vec\nabla p_i|$, 
and that we use $a=1/(2\sqrt{2})$ to exactly counterbalance
solute trapping. 

Only dimensionless parameters remain in the equations, namely,
the already dimensionless coefficients $A_i$, $B_i$, $a_i$, $b$,
and $\tilde\lambda$, and the newly defined $\tilde D = D\bar\tau/W^2$
and 
$\tilde\tau(\vec p)=\tau(\vec p)/\bar\tau$ [with its limiting values
$\tilde\tau_\alpha=\tau_\alpha/\bar\tau$ and 
$\tilde\tau_\beta=\tau_\beta/\bar\tau$], where
$\tau(\vec p)$ is given by Eq.~(\ref{tauofp}). 
The coefficients $A_i(T)$ and 
$B_i(T)$ are just constants for isothermal solidification,
but depend on space and time (through the temperature) for 
directional solidification. Their explicit expression, 
given when choosing the phase diagram in next subsection, 
will contain the rescaled pulling speed 
$\tilde{v_p}=v_p\bar\tau/W$ and thermal lengths 
$\tilde l_{\rm T}^i=l_{\rm T}^i/W$.

We numerically integrate Eqs. (\ref{evolutionofp}) for $p_\alpha$ and
$p_\beta$ only (the equation for $p_{\rm L}$ is redundant) as well as 
Eq.~(\ref{evolutionofmu}); $p_{\rm L}$ is eliminated everywhere through
$p_{\rm L} = 1-p_\alpha-p_\beta$. We use a simple Euler, forward-time, 
centered-space finite-difference scheme with a grid spacing 
$\Delta \tilde x$ and a time step $\Delta \tilde t$ 
slightly below the stability limits of both 
the discretized diffusion and the phase-field equations, the lowest of 
which reads $\Delta \tilde t = (1/4)(\Delta x/W)^2
\min\{1/\tilde D,\tilde\tau_\alpha,\tilde\tau_\beta\}$.
We use standard second-order accurate finite differences, except 
for the Laplacians, which are discretized by a nine-point formula 
involving nearest and next nearest neighbors to reduce lattice anisotropy.
We use $\Delta \tilde x=0.8$ unless otherwise stated, which is the largest 
value for which most results are numerically converged. Exceptionally, 
some parameter sets require $\Delta \tilde x=0.4$.

\begin{figure}
\centerline{\psfig{file=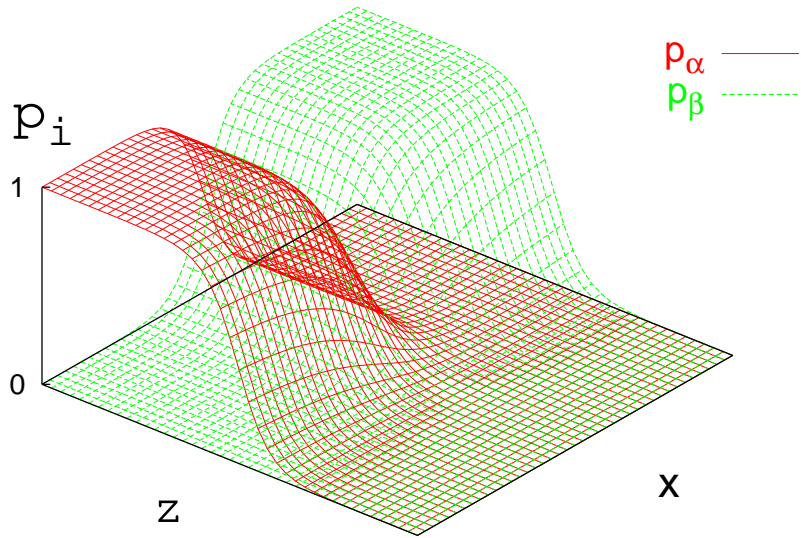,width=.7\textwidth}}
\caption{Illustrative surface plot of the two phase fields $p_\alpha$ and 
$p_\beta$, for two half lamellae as
described in the text. 
$40\times 40$ grid points shown; the system is actually larger in 
the $z$ direction.
}
\label{figpfields}
\end{figure}

We simulate two-dimensional 
directional (or, in one case, isothermal) solidification in a rectangular 
simulation box of total size $n_x \times n_z$, where $n_x$ and $n_z$ are 
the number of grid points in the direction perpendicular and parallel 
to the thermal gradient, respectively.
We consider perfectly periodic lamellar arrays. 
A minimal representation of this geometry
consists of a simulation box with no-flux boundary conditions 
in both the $x$ and $z$ directions 
and two adjacent half lamellae, one of each phase, with their 
centers on the box boundaries and the trijunction in the middle.
Typically, we start with completely flat interfaces, 
and the phase fields are initialized as step functions, 
located at some initial guess for the interface position. 
These step functions then quickly relax to the smooth 
solutions for the phase fields, while the interfaces 
begin to curve and drift to adjust their average undercooling.
Alternatively, the outcome of a previous simulation may also be
taken as initial condition. An example for the resulting configuration
of the phase fields is shown in Fig.~\ref{figpfields}. It can
be clearly seen that the fields smoothly approach their bulk
equilibrium values outside the interfaces, and that,
on each $i$--liquid interface (at $x=$const. sufficiently
far from the solid--solid interface), the other solid phase,
$j\neq i$, is absent ($p_j=0$) through the entire interface 
(as $z$ varies).

In order to reduce the computational effort, we take advantage
of two circumstances. First, as can be seen in Fig.~\ref{figpfields},
most grid points correspond to bulk phases, so that the evolution 
equations can be replaced there by simpler versions, which leads to 
an enormous saving in the total number of operations per time step.
Furthermore, most of the remaining grid points correspond
to purely binary interfaces, so that all the terms in the third, vanishing
phase field can be dropped. Only in the small area near a trijunction
do the full equations need to be integrated.
According to the conditions Eqs. (\ref{valleys}), 
wherever a phase field takes a locally {\em constant}
value of zero (purely binary interface or bulk phase) or unity (bulk phase),
it will remain constant, so that it does not need to be
updated, and its whole evolution equation can be dropped.
In practice, we first compute the Laplacian of each phase field 
everywhere, and then proceed with further calculations for a
phase field $p_i$ only if the modulus of its Laplacian 
exceeds a certain (small) threshold. Similarly, the source and 
antitrapping terms in the evolution equation for the
chemical potential in Eq.~(\ref{evolutionofmu}) 
are evaluated for the locally varying phase field(s) only.
In particular, this means that they can be dropped for bulk phases.
Therefore, in the bulk only the simple diffusion equation needs 
to be solved; in the one-sided model, no field needs to be updated 
in the solid, be it bulk or $\alpha-\beta$ interface.

\begin{figure}
\centerline{\psfig{file=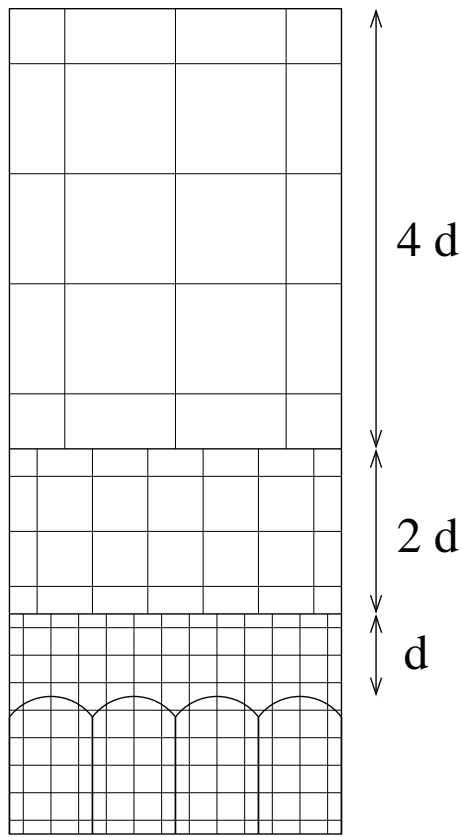,width=.35\textwidth}}
\caption{Configuration of the multi-grid scheme. In front
of the solids that grow upward, a fixed number $n_f$ of
rows is treated on the fine grid with spacing $\Delta x$,
which corresponds to a distance $d=n_f\Delta x$. Two coarser
grids of spacing $2\Delta x$ and $4\Delta x$ have again
$n_f$ rows each, which corresponds to distances $2d$ and
$4d$, respectively. Beyond the coarsest grids, the solution
is one-dimensional.
}
\label{figgrids}
\end{figure}

Second, since the diffusion and thermal lengths are much larger
than the lamellar spacing, the number $n_z$ of grid points required 
in the direction of the thermal gradient is typically much larger 
than $n_x$. A simple analytical calculation (see for example 
Ref.~\cite{jh}) shows that the lateral gradients of the 
diffusion field decay with the distance to the growth front
over a length of the order of the interfacial pattern
($\sim n_x$), whereas in the direction of the pulling the solutal boundary 
layer decays only on the scale of the diffusion length. 
Therefore, the spatial resolution can be decreased with the distance 
to the growth front beyond a scale comparable to $n_x$.
This can be achieved by multi-scale algorithms. We have adapted 
the random walker algorithm of Ref.~\cite{walkers} to eutectic
solidification; however, a disadvantage of this method is its 
numerical noise. For our relatively simple geometry, it is 
straightforward to implement a finite-difference scheme that
uses coarser and coarser grids away from the interface. We
used a hierarchy of in total three grids of increasing length
in the $z$ direction as sketched in Fig.~\ref{figgrids}.
The solutions on the different grids are connected by simple 
linear interpolation. At the end of the coarsest grid, lateral 
concentration gradients are negligibly small; therefore, 
we solve the diffusion equation in one dimension beyond this 
point and connect the solution to the coarsest grid.
We carefully checked that the dependence of the solution on 
the position of the interpolation boundaries is negligible.

Finally, we also follow the solidifying front by advancing the
simulation box whenever the isotherms have advanced by once the
coarsest grid spacing. Since, far enough in the solid, the 
diffusivity vanishes and no evolution takes place, the coldest
part of the solid can be removed without altering the simulation;
new points are added on the liquid side. The boundaries between
grids are also adjusted, and new initial values on the border of the 
finer grids are obtained by interpolation from the coarser grids.

\subsection{Choice of parameters}

The parameters that characterize a given physical situation 
can be grouped into different classes.
First, there are the characteristics of the phase diagram
and some materials parameters that depend only on the alloy system. 
Second, there are the control parameters accessible to the
experimentalist, namely the sample composition and either the
undercooling for isothermal solidification, or the pulling
speed $v_p$ and the temperature gradient $G$ for directional solidification. 
Finally,
there is the lamellar spacing, which cannot be directly
controlled in experiments, but can be fixed in the simulations
by choosing appropriate initial and boundary conditions.

For coupled eutectic growth at low solidification speeds, 
we can make several simplifying assumptions and approximations
in order to keep the number of relevant parameters to
a minimum. Since the average temperature of the solidification 
front is very close to $T_E$, we can (i) use a phase diagram 
linearized around the eutectic point, as specified in 
Eq.~(\ref{gibbsthomson}) of the FBP. This can be implemented 
by setting the parameters $A_i$ and $B_i$ to
\begin{eqnarray}
\label{ai}
A_{\rm L} = 0, \;\; A_i = c_i \mp \frac{(k_i-1)(T-T_{\rm E})}{m_i\Delta C}
            = c_i \mp (k_i-1)
\frac{\tilde z-\tilde{v_p}\tilde t}{\tilde l_{\rm T}^i}, \\
\label{bi} 
B_{\rm L} = 0, \;\; B_i = \mp \frac{A_i (T-T_{\rm E})}{m_i\Delta C}
            = \mp A_i \frac{\tilde z-\tilde{v_p}\tilde t}{\tilde l_{\rm T}^i},
\end{eqnarray}
where the second equalities are obtained using Eq.~(\ref{tempsetup}) 
for the temperature and the definitions of the thermal lengths,
Eq.~(\ref{thermaldef}).
Furthermore, we can
(ii) neglect the relative variation of the concentration gaps 
with respect to their values at $T_E$. This is equivalent to
set $k_i=1$ (parallel liquidus and solidus lines) in 
Eq.~(\ref{mc}) \cite{foot1}. The advantage here is that two 
parameters (the $k_i$) are eliminated and that the $A_i$ become 
independent of temperature.
Finally, at low solidification speeds we can also (iii) neglect 
the kinetic undercooling of the growth front, i.e., we adopt
$\beta_\alpha = \beta_\beta = 0$, which is also the most reasonable 
choice with the information available, since these coefficients are
unknown. All of these approximations were also made 
in the classic Jackson-Hunt analysis \cite{jh}.

Furthermore, we use equal surface tensions for both solid--liquid
interfaces ($\sigma_{\alpha\rm L}=\sigma_{\beta\rm L}$), which
implies equal contact angles $\theta_\alpha=\theta_\beta\equiv \theta$. 
This is a reasonable assumption with the available data for \alloy\,
$\theta_\alpha=70^\circ\pm 10^\circ$ and 
$\theta_\beta=67^\circ\pm 10^\circ$ \cite{mergy}. 
In our test calculations below, we use values of the computational 
parameters for which angles ranging from $\theta=30^\circ$ (all 
surface tensions equal) to $66^\circ$ (case of \alloy) are expected.

With approximations (i) and (ii) above, the phase diagram can be 
characterized by the two ratios
\begin{eqnarray}
\label{r}
r_m & = & m_\beta/m_\alpha, \\
r_c & = & |c_\beta/c_\alpha|.
\end{eqnarray} 
Note that we have $c_\beta-c_\alpha=1$ due to the normalization, and
hence it follows that $c_\beta=r_c/(r_c+1)$ and $c_\alpha=-1/(r_c+1)$.
The measured properties for \alloy\ are listed in Table~\ref{paramtable}.
According to the constraint of Eq.~(\ref{thermoconstraintatte}), for
equal surface tensions we should have $d_\alpha/d_\beta=r_c$.
From Table~\ref{paramtable}, we find $d_\alpha/d_\beta = 2.7\pm 0.5$, 
while $r_c = 2.5\pm 0.5$. Thus, the constraint 
is respected to within the accuracy of the measured 
data. For our simulations, we have to choose a set of values that
exactly satisfies the constraint and is compatible with the
experimental error bars. We use the concentrations to fix 
$r_c=d_\alpha/d_\beta=c_\beta/c_\alpha=2.5$, and take as 
a reference length the average
\begin{equation}
\bar d = (d_\alpha + d_\beta)/2
\end{equation}
of the measured capillary lengths, $\bar d=6.5$nm. The
capillary lengths obtained using this value of $\bar d$
together with $r_c=2.5$ and the constraint are listed
on the last column of Table~\ref{paramtable}. Similarly,
we choose $r_m=2$ and fix a corresponding pair of values
for the liquidus slopes within the error bars.
Finally, we adopt the mean measured value
of the solute diffusivity, $D=0.5\times 10^{-9}$ m$^2$/s.
The measured and used values of the parameters 
are summarized in Table~\ref{paramtable}.

%
\begin{table}
\caption{Materials parameters of the transparent alloy \alloy,
as measured in Ref.~\protect\cite{mergy}, and the values used 
for the present study (see text). The $d_i$ listed under Ref.~\cite{mergy} 
were obtained using the basic definitions of Eq.~(\ref{capillarydef}) 
together with the thermophysical data from that reference.
The value of $C_{\rm E}$ used corresponds to $r_c\equiv 2.5$.
\label{paramtable}}
\begin{tabular}{lccc}
Quantity & Symbol & Ref.~\protect\cite{mergy} & Value used \\
\hline
Liquidus slope of $\alpha$ & $m_\alpha$ & $-(81\pm 5)$ K/mol & -82 K/mol \\
Liquidus slope of $\beta$ & $m_\beta$ & $ (165\pm 5)$ K/mol & 164 K/mol \\
Composition of $\alpha$ at $T_E$ & $C_\alpha$ & $(8.8\pm 0.4)$ mol\% & 8.8 mol\% \\
Composition of $\beta$ at $T_E$ & $C_\beta$ & $(18.5\pm 0.9)$ mol\% & 18.5 mol\% \\
Eutectic composition & $C_E$ & $(11.6\pm 0.6)$ mol\% & 11.571 mol\% \\
Capillary length of $\alpha$ & $d_\alpha$ & $(9.5\pm 2.0)$ nm & 9.29 nm \\
Capillary length of $\beta$ & $d_\beta$ & $(3.5\pm 1.0)$ nm & 3.71 nm \\
Contact angle of $\alpha$ & $\theta_\alpha$ & $(70\pm10)^\circ$ & 30$^\circ$--66$^\circ$ \\
Contact angle of $\beta$ & $\theta_\beta$ & $(67\pm10)^\circ$ & 30$^\circ$--66$^\circ$ \\
Partition coefficient of $\alpha$ & $k_\alpha$ & 0.75 & 1.0 \\
Partition coefficient of $\beta$ & $k_\beta$ & 1.5 & 1.0 \\
Impurity diffusivity & D & $(0.5\pm 0.1)\times 10^{-9}$ m$^2$/s\ \  & $0.5\times 10^{-9}$ m$^2$/s\\
\end{tabular}
\end{table}

The control parameters which can be varied in an experiment
enter our simulations as follows.
The global sample composition is fixed by imposing the appropriate
value for the chemical potential $\mu(z\to\infty)=c_\infty$ 
(recall that $A_{\rm L}=0$) as a boundary condition on the
liquid side of the simulation box. For isothermal
solidification, the undercooling enters through Eqs. (\ref{ai})
and (\ref{bi}) in dimensionless form, $\Delta_i=(T_{\rm E}-T)/(m_i\Delta C)$.
For directional solidification, the pulling
speed $v_p$ and the thermal gradient $G$ enter through 
the same equations {\it via} 
the diffusion and thermal lengths $l_D=D/v_p$ and $l_T^i=m_i\Delta C/G$. 
Note that $l_T^\beta/l_T^\alpha = r_m$,
so that $l_T^\alpha=[2/(1+r_m)]\bar l_T$ and 
$l_T^\beta=[2r_m/(1+r_m)]\bar l_T$,
where we recall that $\bar l_T=(l_T^\alpha+l_T^\beta)/2$
is the average thermal length. Finally, the lamellar
spacing $\lambda$ (not to be confused with the coupling
constant $\tilde\lambda$) is fixed by the lateral size 
of the simulation box.

To summarize, the physical conditions to simulate are
completely specified by the two asymmetry parameters
$r_c$ and $r_m$ and four lengths, namely, $\bar d$, $\bar l_T$,
$l_D$ and $\lambda$. From these four length scales, three dimensionless
parameters can be constructed. We choose $l_D/\bar d$, 
$\lambda/\lambda_{\rm JH}$, and $\bar l_T/l_D$. The
second choice is motivated by the fact that the Jackson-Hunt
minimum undercooling spacing is a reference length for
eutectic pattern formation. Dividing both sides of
Eq.~(\ref{jhlambda}) by $\bar d$ yields 
$\lambda_{\rm JH}/\bar d \propto \sqrt{l_D/\bar d}$, where
the proportionality constant depends only on the sample
composition (through the volume fraction $\eta$) 
and on $r_c$ (through the ratios $d_i/\bar d$).
Therefore, specifying $\lambda/\lambda_{\rm JH}$ and $l_D/\bar d$
fixes implicitly $\lambda/\bar d$.

All physical conditions fixed, we now choose the only truly free
computational parameter, the interface thickness $W$. The
relevant scale of the pattern is of course 
the lamellar spacing $\lambda$, and
therefore the resolution of the phase-field simulations
is given by the ratio $\lambda/W$. Let us first outline
the procedure to determine the parameters when the spacing
is given in physical units (meters). Then, $\lambda/W$
directly fixes $W$ in meters. Next, the coupling constant
$\tilde\lambda$ is determined from Eqs.~(\ref{capillary})
for the capillary lengths in terms of phase-field
parameters (recall that $A_{\rm L}=0$),
\begin{equation}
\tilde \lambda = \frac{W}{\bar d} \frac{a_1}{2} 
   \left(\frac{1}{|A_\alpha|}+\frac{1}{|A_\beta|}\right).
\label{w}
\end{equation}
Finally, the relaxation times $\tau_i$ (in seconds) are
obtained (for arbitrary kinetics) using Eq.~(\ref{kinetic}),
\begin{equation}
\tau_i = \tilde \lambda |A_i| W
   \left(\frac{\beta_i}{a_1} + a_2\frac{|A_i|W}{D}\right),
\label{taus}
\end{equation}
which yields $\bar\tau=(\tau_\alpha+\tau_\beta)/2$ and the
$\tilde\tau_i=\tau_i/\bar\tau$. The diffusion coefficient
$\tilde D$, pulling speed $\tilde v_p$, and thermal lengths
$\tilde l_{\rm T}^i$ needed in Eqs.~(\ref{evolutionofp}) 
and (\ref{evolutionofmu}) are then obtained by scaling 
the corresponding dimensional quantities by $W$ and $\bar\tau$.
For isothermal solidification, the dimensionless undercooling
$\Delta_i$ is directly obtained from the actual temperature
$T$ and the alloy properties $T_{\rm E}$, $m_i$ and $\Delta C$.

An alternative way to obtain the simulation parameters
is to start directly from the dimensionless ratios. Indeed,
specifying $\lambda / W$ for fixed $\lambda/\bar d$ fixes
directly $W/\bar d$ and hence $\tilde\lambda$ from Eq.~(\ref{w}).
For $\beta_\alpha=\beta_\beta=0$, Eq.~(\ref{taus}) yields
$\tilde\tau_\beta/\tilde\tau_\alpha=(|A_\beta|/|A_\alpha|)^2$.
For the case $k_i=1$ considered here, $A_i=c_i$, and hence
$|A_\beta|/|A_\alpha|=r_c$, and we obtain directly 
$\tilde\tau_\alpha=2/(1+r_c^2)$ and $\tilde\tau_\beta=2r_c^2/(1+r_c^2)$. 
With these values fixed, Eq.~(\ref{taus})
with the $c_i$ given after Eqs. (\ref{r}) 
yields the scaled diffusivity,
\be
\label{dofbutton}
\tilde D = (1/2) a_2\tilde\lambda (1+r_c^2)/(1+r_c)^2.
\ee
Since $W/\bar d$ and $l_D/\bar d$ are both known, the
scaled pulling speed can then be inferred 
from $l_D/W=\tilde D/\tilde v_p$; finally, the scaled thermal
lengths are obtained via the ratio $\bar l_T/l_D$ for directional
solidification, or the dimensionless undercooling $\Delta_i$ is directly
plugged into the $A_i$'s and $B_i$'s for isothermal solidification.

To illustrate the above procedure, we show in Tables \ref{symmtable} 
and \ref{cbrtable} the computational parameters for different 
values of the interface thickness for two series of test 
simulations that will be discussed in detail below.
Both sets of simulations are carried out at the eutectic
composition and for a spacing $\lambda=\lambda_{\rm JH}$.
The first is for a model alloy with a symmetric phase
diagram, $r_m=r_c=1$. For such an alloy at the eutectic 
composition, all equations are completely symmetric with 
respect to the interchange of the two solid phases. In
this case, we use $\bar{l}_T/l_D=4$ and $l_D/\bar d=51200$.
The second set of simulations is for the
phase diagram of \alloy, $r_c=2.5$ and $r_m=2$, and we use
$l_D/\bar d = 41796$ and $\bar l_T/l_D=4$, which corresponds
to a pulling speed of $v_p\approx 1.8$ $\mu$m/s and a temperature 
gradient of $G\approx 110 K/$cm, both fairly typical experimental
values.

\begin{table}
\caption{Simulation parameters used in the runs for the symmetric 
model alloy. Physical parameters are $l_D/\bar d = 51200$, 
$\bar l_T/l_D = 4$, $\lambda=\lambda_{{\rm JH}}$, and the sample is at 
eutectic composition. Space and time units are $W$ and $\bar\tau$, 
respectively.
\label{symmtable}}
\begin{tabular}{cccccc}
$\lambda/W$ & $\tilde D$ & $\tilde\lambda$ & $\tilde v_p$ & $\bar l_T/W$ & $\Delta t/\bar\tau$ \\
\hline
32 & 10.633 & 36.197 & 0.0079729 & 5334.5 & 0.012038 \\
64 & 5.3164 & 18.098 & 0.0019932 & 10669 & 0.024077 \\
96 & 3.5442 & 12.066 & 0.00088588 & 16003 & 0.036115 \\
128 & 2.6582 & 9.0491 & 0.00049831 & 21338 & 0.048153 \\
\end{tabular}
\end{table}

\begin{table}
\caption{Simulation parameters used in the runs for \alloy. 
Physical parameters are $l_D/\bar d = 41796$, $\bar l_T/l_D = 4$, 
$\lambda=\lambda_{{\rm JH}}$, and the sample is at eutectic composition. 
Space and time units are $W$ and $\bar\tau$, respectively.
\label{cbrtable}}
\begin{tabular}{ccccccc}
$\lambda/W$ & $\tilde D$ & $\tilde\lambda$ & $\tilde v_p$ & $l_T^\alpha/W$ & $l_T^\beta/W$ & $\Delta t/\bar\tau$ \\
\hline
32 & 14.852 & 42.713 & 0.013141 & 3013.7 & 6027.4 & 0.0086186 \\
64 & 7.4258 & 21.357 & 0.0032854 & 6027.4 & 12055 & 0.017237 \\
96 & 4.9505 & 14.238 & 0.0014602 & 9041.1 & 18082 & 0.025856 \\
128 & 3.7129 & 10.678 & 0.00082134 & 12055 & 24110 & 0.034474 \\
\end{tabular}
\end{table}

Finally, the $a_i$ are dimensionless parameters that control
the ratio of surface tensions through the relative
heights of the free-energy barriers between bulk phases.
They hence provide a handle on the contact angles.
The choice of the $a_i$ is independent of all the other
parameters.
For equal solid--liquid surface tensions, we recall
that $a_\alpha=a_\beta=0$;
$a_{\rm L}$ will then be varied in the next subsection
to tune the ratio of solid--liquid to solid--solid surface tensions,
and thus change the solid--liquid contact angle $\theta$.
A change from $a_{\rm L}=0$ to $a_{\rm L}=12$ is necessary
to span contact angles from 0$^\circ$ to 66$^\circ$ as desired. 
On the other hand, $b$ needs to be tuned to ensure the convexity
of the free-energy landscape; we test
values ranging from $b=3$ to $b=12$.

\subsection{Isothermal solidification and contact angles}

We begin our simulations by testing whether our model reproduces 
the correct contact angles at the trijunction point.
For that purpose, we need to 
extract first the interface shapes, then the trijunction 
position, and finally the angles between interfaces from the simulations.
We proceed as follows.
 
The difference $p_j-p_i$ is computed everywhere, 
for every pair of phases $i\neq j$. Wherever one of these combinations 
changes sign, we interpolate the position of a $i$--$j$ interface point 
from the two adjacent values of $p_j-p_i$; we also interpolate the 
value of the third phase $p_k$ at that point. If $p_k<1/3$, the
point lies on a ``true'' $i$--$j$ interface; otherwise, it is located 
inside the third bulk phase on the ``prolongation'' of the interface 
beyond the trijunction point. In most cases, the latter points are 
not plotted, and the $i$--$j$ interface terminates at the trijunction; 
however, as we will see below, it is sometimes useful to plot the 
prolongation as well, because it can yield information about the 
internal structure of the trijunction. The average undercooling of 
each interface is obtained from the $z$ position of all of its 
points. In order to avoid sharp cutoffs and discretization errors 
at trijunctions, the contribution of each point on the $p_i=p_j$
isocontour to the undercooling is weighted by $1-p_k$, rather 
than simply counting points on the ``true'' interface and discarding 
points on the prolongation.

When all three interfaces enter and leave a particular 
elementary grid square exactly once, 
their entrance and exit positions computed
by the method described above define one straight segment for each interface.
If all three segments intersect with each of the other two, a triple point 
is considered to be detected at the average position of the three intersections, 
provided that this lies within the grid square. While this procedure yields
an excellent subgrid resolution for the position of the trijunction points, 
the determination of the angles from the slopes of the segments is 
subject to large grid effects (that is, the values obtained are found
to depend on the position of the trijunction with respect to the grid
points). Therefore, 
we use a more precise procedure: Around a detected trijunction point, 
a few interface points on both side of the trijunction are first obtained 
with high precision using nonlinear interpolations. Then, two such points on 
each side of the trijunction are used together with a nonlinear 
interpolation to obtain the angles at the trijunction position. With 
this procedure, the uncertainty on the angles due to grid effects is 
reduced to about $0.1^\circ$.

Before engaging in directional solidification,
we test both the model and the above procedures by
simulations of isothermal solidification, in which the temperature
is set to a constant value below the eutectic temperature; in the
free-boundary problem, 
the terms $(z_{\rm int}-v_pt)/l_T^i$ are then replaced by 
constants $-\Delta_i$ where $\Delta_i=(T_{\rm E}-T)/(m_i \Delta C)$ 
is the dimensionless undercooling of phase $i$ ($i=\alpha,\beta$).
Since the interface temperature is fixed, the quantity to be 
selected is now the interface velocity. This velocity is negative 
for low undercooling (the lamellae melt), positive for high
undercooling, and exactly zero for a critical undercooling that
depends on the lamellar spacing through the curvature of the
solid--liquid interfaces. For the symmetric model alloy at the
eutectic composition, an exact solution for this critical steady
state is known: Since the composition in the liquid and the
temperature are uniform and the velocity is zero, the interface
curvature is constant [see, for instance, Eq. (\ref{gibbsthomson})], 
so that the two solid--liquid interfaces
form circular arcs that intersect at the trijunction point. For
a given contact angle and lamellar spacing, the critical undercooling is
\be
\label{deltacrit}
\Delta_c = 4\sin\theta\frac{\bar d}{\lambda}.
\ee

We conducted series of simulations with different undercoolings
and extracted the velocity of the solid--liquid interface once
it had reached a steady state; the critical undercooling was 
obtained by seeking the zero crossing of the velocity. The 
grid spacing was fixed to $\Delta x=0.4$ to better resolve the
neighborhood of the critical undercooling (small velocities), 
and the parameters of the first line in Table~\ref{symmtable}
(except $\tilde{v}_p$ and $\tilde{l}_T$) were used. 
We tested five different values of $a_{\rm L}$
for which equilibrium angles from $\theta=30^\circ$ to $\theta=66^\circ$ 
are expected from Young's law, with a constant $b=12$. 
The predicted angle was then plugged into Eq. 
(\ref{deltacrit}) to obtain the theoretical value of the 
critical undercooling. The latter is then compared to its measured
value, as summarized in 
Table~\ref{undertable}. We find excellent agreement
between simulations and theory; the error 
in the undercooling increases with $a_{\rm L}$
(and hence with the contact angle), but remains of the
order of $1$\% even for $a_{\rm L}=12$, which corresponds to
$\theta=66^\circ$.
In particular, this implies that Young's law is satisfied.
%
\begin{table}
\caption{Critical undercooling for stationary lamellar states for
various values of $a_{\rm L}$ (and hence of the contact angle $\theta$);
$b=12$ in all cases, and the other parameters are given in the 
first line of Table \protect\ref{symmtable}.
\label{undertable}}
\begin{tabular}{ccccc}
$a_{\rm L}$ & $\theta$ expected & $\Delta_c$ (simulation) 
& $\Delta_c(\theta)$ (theory) & error \\
\hline
0 & 30$^\circ$ & 0.001624 & 0.001628 & 0.25\% \\
3 & 39$^\circ$ & 0.002047 & 0.002055 & 0.39\% \\
6 & 48$^\circ$ & 0.002390 & 0.002408 & 0.75\% \\
9 & 56$^\circ$ & 0.002686 & 0.002713 & 1.00\% \\
12 & 66$^\circ$ & 0.002946 & 0.002987 & 1.37\% \\
\end{tabular}
\end{table}
%

However, if we measure the contact
angles {\em directly} using the numerical procedure outlined above, we find a 
consistent result only for $a_{\rm L}=0$, that is, the measured contact 
angles are {\em always} very close to $30^\circ$; 
for $a_{\rm L}\neq 0$, we find 
angles that differ widely from the expected values. Furthermore, 
whereas the detected critical undercooling is almost independent 
of the parameter $b$, the measured angles for 
$a_{\rm L}=12$ 
vary by $20^\circ$ when $b$ is varied between $3$ and $12$.

\begin{figure}
\centerline{\psfig{file=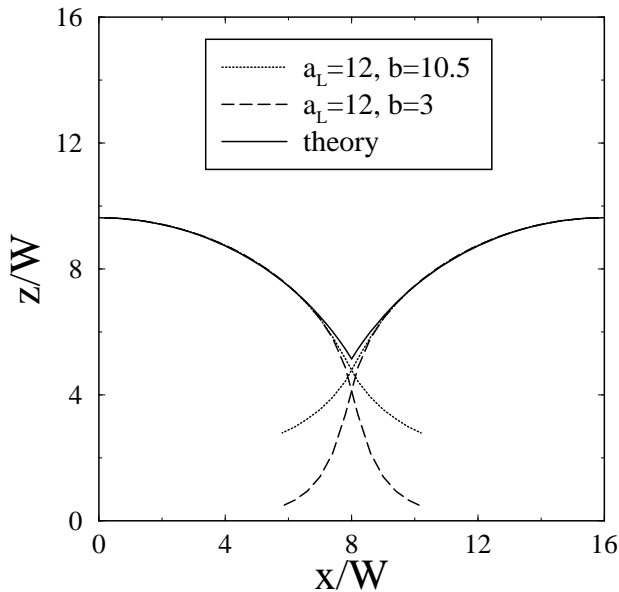,width=.45\textwidth}}
\caption{Lamellar states, simulated with a constant temperature
that yields almost zero growth speed, and comparison to the
analytical critical stationary solution that consists of circular arcs.
Two different values of $b$ are used together with $a_{\rm L}=12$
(predicted angle $\theta\approx 66^\circ$). Since, at constant temperature,
the origin of the z axis is arbitrary, the solutions have
been shifted in order for lamella tips to coincide. The
curves shown are the isocontours $p_\alpha=p_{\rm L}$ and $p_\beta=p_{\rm L}$,
to be compared with the theoretical circle arcs
with the predicted contact angle $\theta$ (solid curve).
The isocontour $p_\alpha=p_\beta$, exactly vertical and located at 
$x/W=8$, has been omitted for clarity.
}
\label{figtristruc}
\end{figure}

The reason for this behavior becomes apparent in Fig.~\ref{figtristruc}.
We show the analytical solution, which consists of two circular arcs
with contact angles $\theta=66^\circ$,
as well as the interfaces extracted from the simulations for two
different values of $b$. Whereas the agreement between the two
simulations and the theory is excellent far from the trijunction,
the three shapes start to become different at a distance of about 
$W$ from the trijunction. The lines $p_i=p_j$ exhibit curvatures
which differ from the curvature of the circular arcs of the analytical
solution, and which depend on the value of $b$; the position of
the trijunction point also depends on $b$. This behavior is due 
to the changes in the potential energy landscape induced by the
variations of $b$, which affects the internal structure of the
trijunction region. Note, however, that the radius of curvature
of the solid--liquid interfaces far from the trijunction does not
change. This implies that the critical undercooling remains the
same, and hence that Young's condition and the contact angles
extrapolated from the interfaces to the trijunction are correctly
implemented in all cases.

The lesson from all this is that the local procedure outlined
above does in general not yield the ``macroscopic'' angles,
but values that are influenced by the internal structure
of the trijunction region. A different possibility to obtain
the ``macroscopic'' angles would be to construct the
intersection of the two solid--liquid interfaces, extrapolated
from their shape outside the trijunction. However, this
procedure is prone to large errors out of equilibrium, since 
the curvature varies along the interfaces.  
As a consequence, contact angles are very hard to measure in
out-of-equilibrium situations when $\theta\neq 30^\circ$.
When $\theta=30^\circ$, the ``local'' procedure 
works properly.
The particularity of this value is, of course, the symmetry 
between all three phases, which entails that the prolongation 
of the interface inside the third phase runs exactly in 
the middle between the two other interfaces and is not 
``deflected'' from the direction in which it enters the 
trijunction. In contrast, for $a_{\rm L}\neq 0$, the local
shape and curvature of this isoline depend on the internal
structure of the trijunction, which explains the errors
made in the ``local'' measurement of the angles.

Since we are interested in precise information on the
angles, we restrict all the following simulations to the
case of equal surface tensions, where reasonably accurate
measurements of the angles can be carried out. An additional
advantage of this choice is that it avoids the reduced
grid spacing needed to resolve the steeper interfaces
associated to a higher surface tension. 
For $a_{\rm L}=12$ (which corresponds to the actual value of
$\theta$ in \alloy), $\Delta x$ needs to be divided by a factor
of 2 ($\Delta x=0.4$) with respect to $a_{\rm L}=0$, which implies 
using a 4 times larger system size (in terms of grid points)
running for 4 times more time steps (recall that 
$\Delta t\propto \Delta x^2$), which hence takes 16 times 
more CPU time.
Having tested the equilibrium angles, in the remaining
subsections we focus on out-of-equilibrium simulations
of directional solidification.

\subsection{Steady state lamellae}
The convergence of our model to the thin-interface predictions
is tested by performing series of simulations for fixed physical 
parameters and decreasing interface width. For each simulation, 
the interface profiles and interface undercooling are monitored 
to check when the steady state is reached. For comparison, the 
same series of runs is also performed without the antitrapping 
current, and for an earlier version of our model presented 
in Ref.~\cite{dresden} that uses different interpolation functions $g_i$. 
Since those functions are not antisymmetric with respect to the point
$p_i=p_j=1/2$ on $i$--$j$ interfaces [i.e., they do not satisfy
Eq. (\ref{gsymmetry0})], this model exhibits several 
thin-interface corrections \cite{onesided,almgren}.

\begin{figure}
\hbox{
{\vbox{\psfig{file=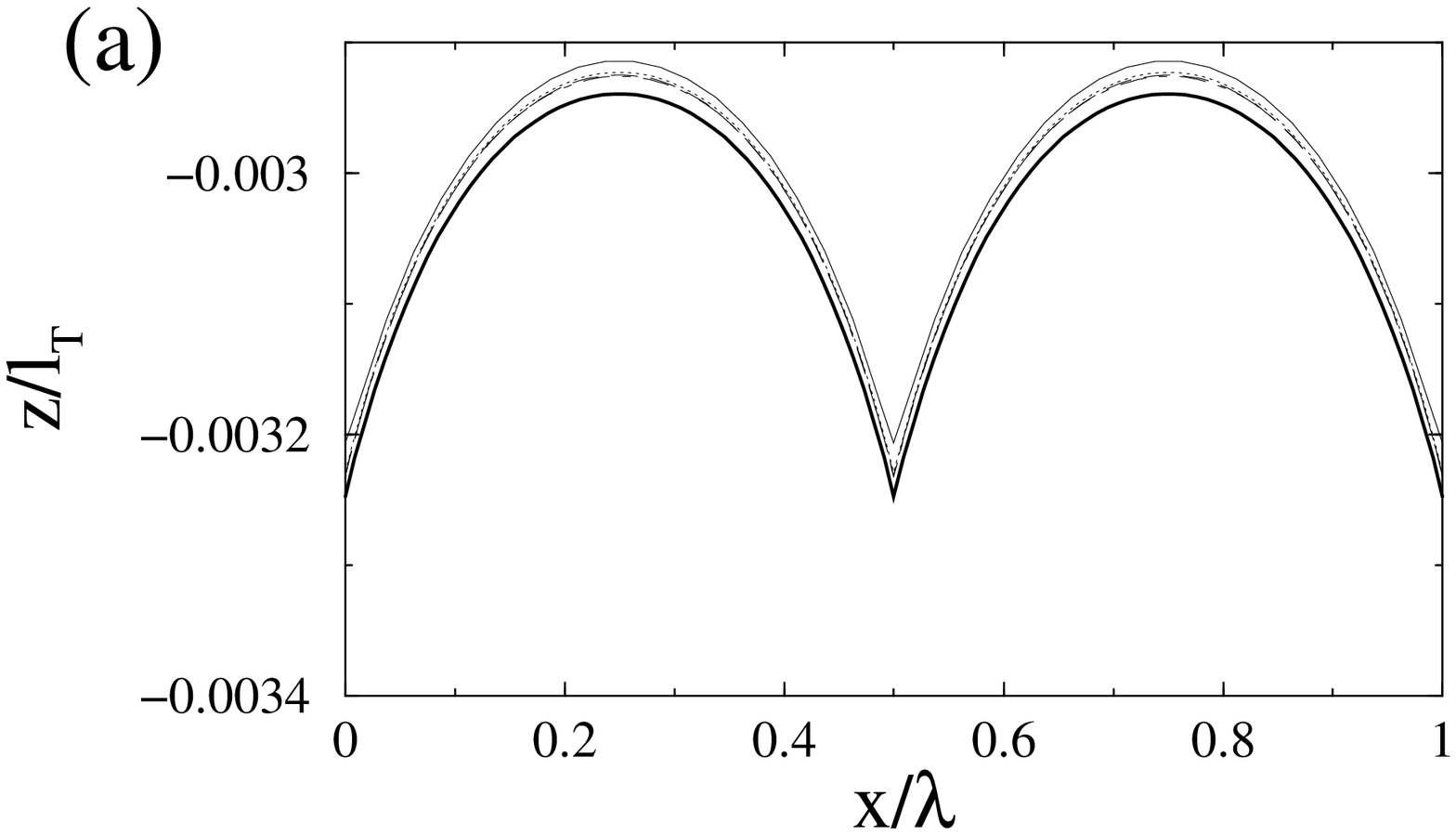,width=8.2cm}
\psfig{file=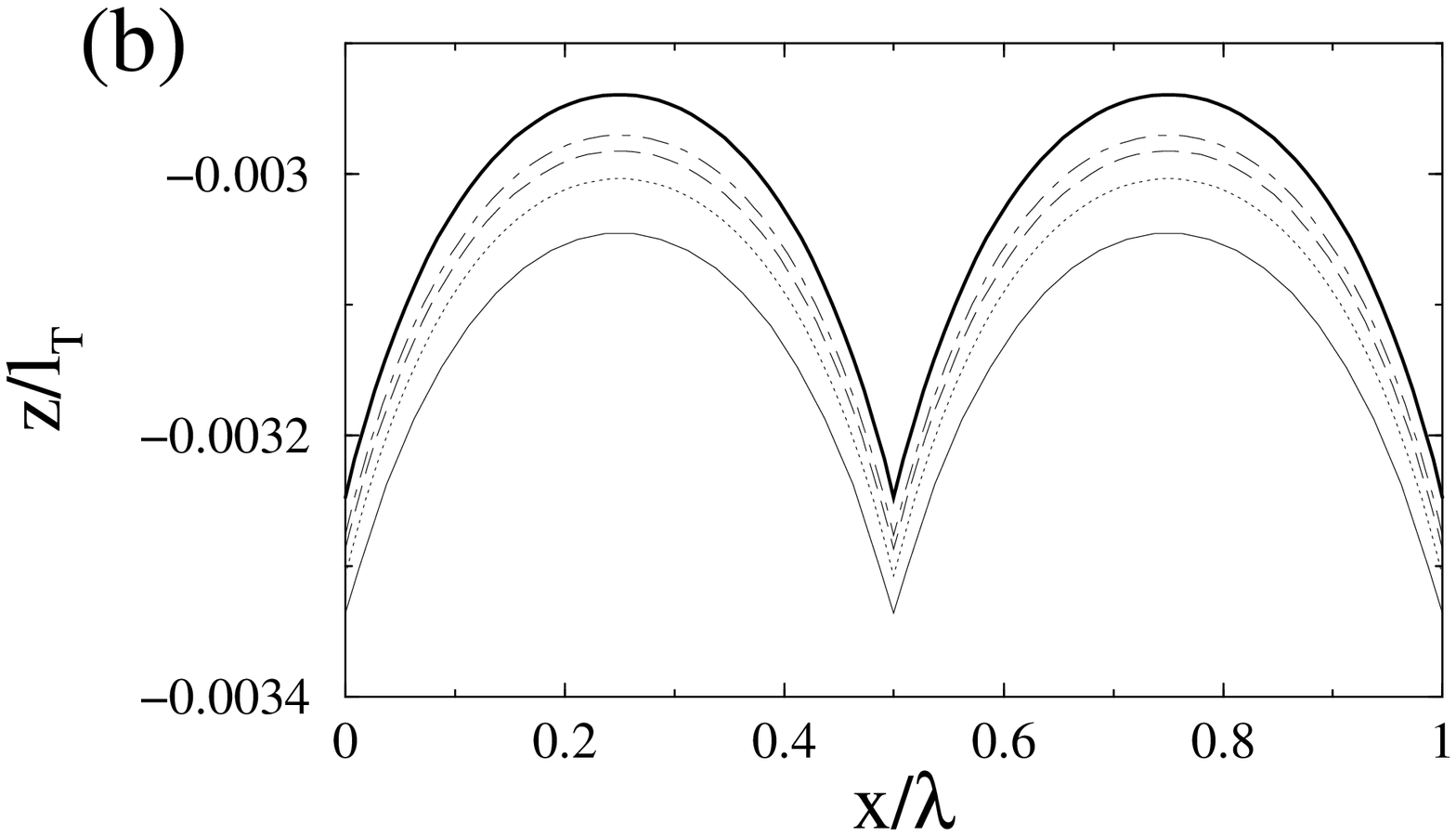,width=8.2cm}}
\hskip 9mm
\vbox{ 
\psfig{file=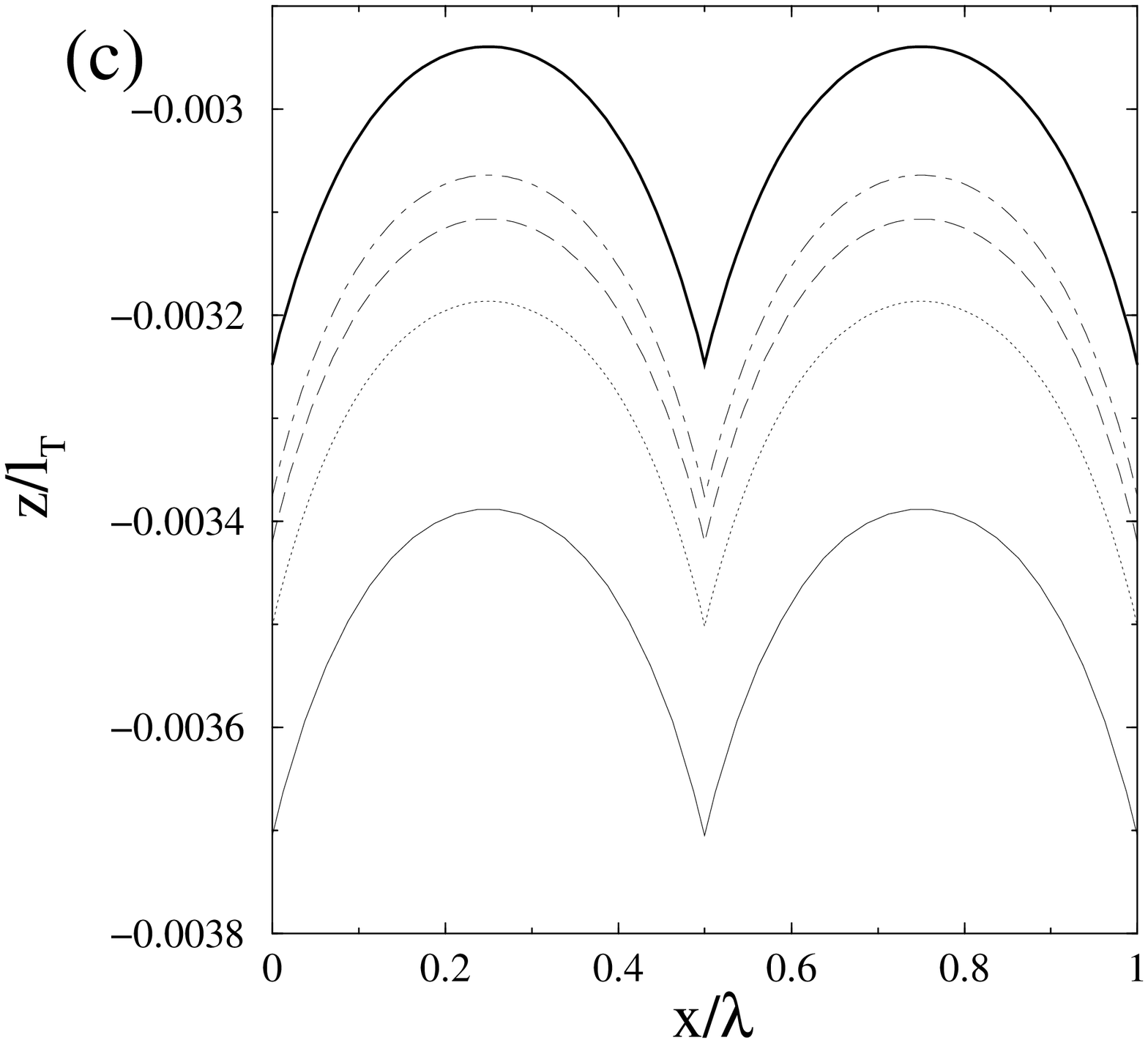,width=8.2cm}
$\;$}}}
\caption{Convergence test for lamellar shapes in the symmetric
model alloy at the eutectic composition. The solid--liquid
interfaces are plotted for (a) the complete model, (b)
without the antitrapping current, and (c) a model with several
thin-interface corrections taken from Ref.~\protect\cite{dresden}. 
The parameters for all runs are given in Table~\protect\ref{symmtable}.
Thin solid lines: $\lambda/W=32$, dotted lines: $\lambda/W=64$,
dashed lines: $\lambda/W=96$, dash-dotted lines: $\lambda/W=128$;
thick solid line: result of the boundary integral code of 
Ref.~\protect\cite{Karma96}.
}
\label{figsymm}
\end{figure}

The parameters used for the simulations of the model alloy with symmetric
phase diagram are listed in Table \ref{symmtable}. Simulation times on
a 2.4 GHz Intel Xeon processor range from half an hour for the lowest 
resolution ($\lambda/W=32$) to three days for the highest ($\lambda/W=128$).
The results for the interface profile are shown in Fig.~\ref{figsymm},
together with results of boundary integral calculations performed
with the code of Ref.~\cite{Karma96}.
Whereas for the present model all curves for $\lambda/W \ge 64$
superimpose perfectly, for the other models large errors appear.
It can be seen that they are smaller for the model that only lacks
the antitrapping current; however, a decisive progress is made
only when all thin-interface corrections are eliminated. Even so,
a small difference with the boundary integral prediction
remains in the $W/\lambda \to 0$ limit. 
However, a close examination reveals that the two solutions 
are simply shifted with respect to each other, with is due either 
to a residual interface kinetics in the phase-field model or to the
approximations of the boundary integral method. The relative error
in the average undercooling is about 0.3 \%.
Since the advantage of the complete model over the other
formulations is obvious, it has been used exclusively for
all the remaining simulations.

\begin{figure}
\centerline{\psfig{file=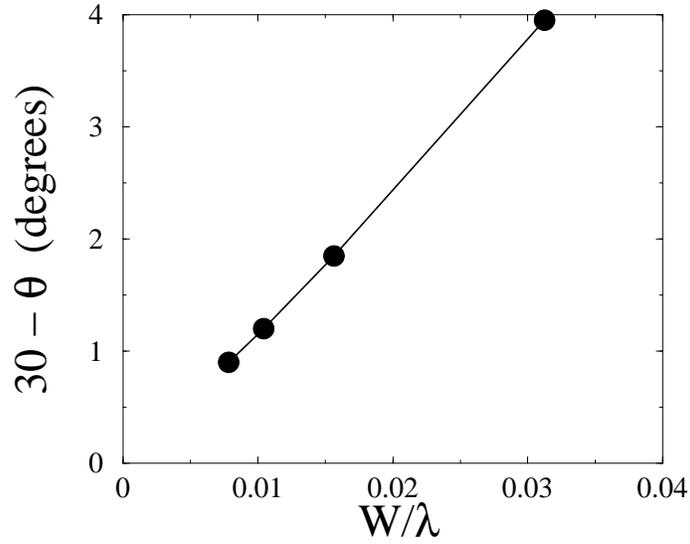,width=.5\textwidth}}
\caption{Difference between the contact angle and the equilibrium
value $30^\circ$ as a function of $W/\lambda$ for the same series
of runs as Fig.~\protect\ref{figsymm}(a).
}
\label{figangle}
\end{figure}

We also examine the contact angles at the trijunction point for
the same series of runs with the complete model.
We find small deviations from the
expected equilibrium value of $30^\circ$. In Fig.~\ref{figangle},
we plot the difference, $30^\circ - \theta$, as a function of
$W/\lambda$. It can be seen that it extrapolates to zero in
the $W/\lambda\to 0$ limit. This indicates that there is
a finite-interface-thickness correction to the 
{\em non-equilibrium}
contact angles
that vanishes in the sharp-interface limit. Since the difference
is smaller than $2^\circ$ for $\lambda/W \ge 64$, this effect is
invisible in the profiles and does not appreciably influence
the results for the undercooling.
In particular, it cannot be responsible for the 0.1\% undercooling
mismatch described above, since the latter does not extrapolate
to zero in the $W/\lambda \to 0$ limit.

\begin{figure}
\centerline{\psfig{file=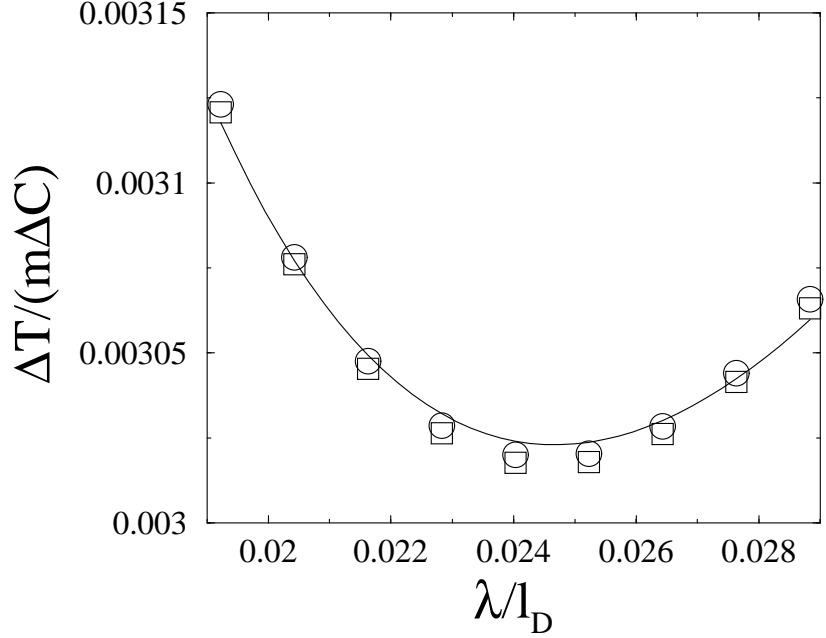,width=.6\textwidth}}
\caption{Dimensionless average undercooling of the front
versus dimensionless spacing ($l_D=D/v_p$ is the diffusion length)
for the first two parameter sets of Table~\protect\ref{symmtable}. 
The curve is a best fit to the Jackson-Hunt law (see text for details).
}
\label{figJH}
\end{figure}

We now consider the dependence of the average undercooling of
the front $\Delta T$ on the lamellar spacing $\lambda$, 
which is known as the Jackson--Hunt curve,
Eq.~(\ref{jhglobal}). We perform a series of runs for the
same materials and computational parameters as above, but with 
varying box size (and hence lamellar spacing). For the resolution, 
we take $\lambda_{\rm JH}/W=32$ and 64, since, on the basis of
Fig. \ref{figsymm}(a),
we expect results to be converged for $\lambda_{\rm JH}/W\ge 64$.    
The curves thus obtained are plotted in Fig.~\ref{figJH},
for the first ($\lambda_{\rm JH}/W=32$, squares) and second 
($\lambda_{\rm JH}/W=64$, circles) set of
parameters of Table~\ref{symmtable}. The undercooling $\Delta T$,
scaled by $m\Delta C$, can be directly obtained through
$\Delta T/(m\Delta C) = \langle z_{\rm int}\rangle/l_T$
(note that for the symmetric model alloy, the two liquidus 
slopes and hence also the two thermal lengths are equal, 
$m_\alpha=m\beta=m$ and $l_T^\alpha=l_T^\beta=l_T$),
where $\langle z_{\rm int}\rangle$ is the $z$ position
of the solid-liquid interface, averaged over the lateral
coordinate $x$. The line shows the best 
fit of the higher resolution data to the Jackson-Hunt law, 
Eq.~(\ref{jhglobal}), with $\Delta T_{{\rm JH}}$ and $\lambda_{{\rm JH}}$ 
as free parameters. The fit yields $\lambda_{{\rm JH}}/l_D=0.02468$ and
$\Delta T_{{\rm JH}}/(m\Delta C)=0.003023$, whereas the theoretical
values calculated from Eqs.~(\ref{jhlambda}) and (\ref{jhdeltat})
are $0.02403$ and $0.003251$, respectively; this corresponds
to relative differences of 2.7 \% and 7.0 \%, respectively.
This good agreement is especially noteworthy because
in phase-field models which exhibit thin-interface corrections
the difference between simulated and calculated minimum
undercooling spacings is usually much 
larger \cite{Karma94,Plapp02,Kim04}.
Note that the Jackson--Hunt prediction for the
minimal undercooling and its corresponding spacing is based on 
an approximate description of the front. We know from 
Fig.~\ref{figsymm}(a) that the difference
in average undercooling between the phase-field model and 
the boundary-integral method is only 0.3 \% near the minimum undercooling.
Therefore, the 7 \% difference with the Jackson--Hunt value must
be attributed to the approximations in the theory, and not
to the phase-field model.

\begin{figure}
\centerline{\psfig{file=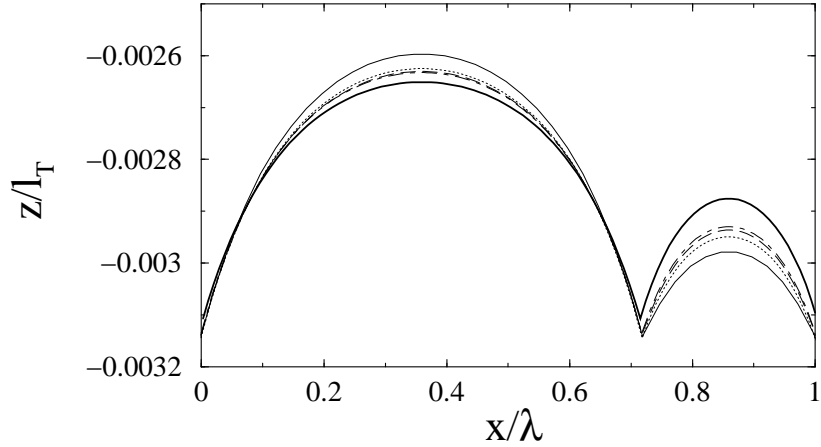,width=.6\textwidth}}
\caption{Convergence test for lamellar shapes in \alloy. The
parameters of all runs are given in Table~\protect\ref{cbrtable}.
Thin solid lines: $\lambda/W=32$, dotted lines: $\lambda/W=64$,
dashed lines: $\lambda/W=96$, dash-dotted lines: $\lambda/W=128$;
thick solid line: result of the boundary integral code of 
Ref.~\protect\cite{Karma96}.
}
\label{figcbr}
\end{figure}

Next, we repeat a similar procedure for the phase diagram
of \alloy. The parameters are listed in Table.~\ref{cbrtable},
and the resulting interface shapes shown in Fig.~\ref{figcbr}.
The behavior is qualitatively similar, but the convergence is
slower. It can be seen that successive interface shapes fall
closer and closer together as the resolution is increased. For
the larger $\alpha$ lamella, the shape is converged for 
$\lambda/W \ge 64$; for the smaller $\beta$ lamella, tiny differences
remain visible even between $96$ and $128$. This is not surprising:
convergence is harder to achieve when smaller physical details (such
as the small $\beta$ lamella here) must be resolved.

\begin{figure}
\centerline{\psfig{file=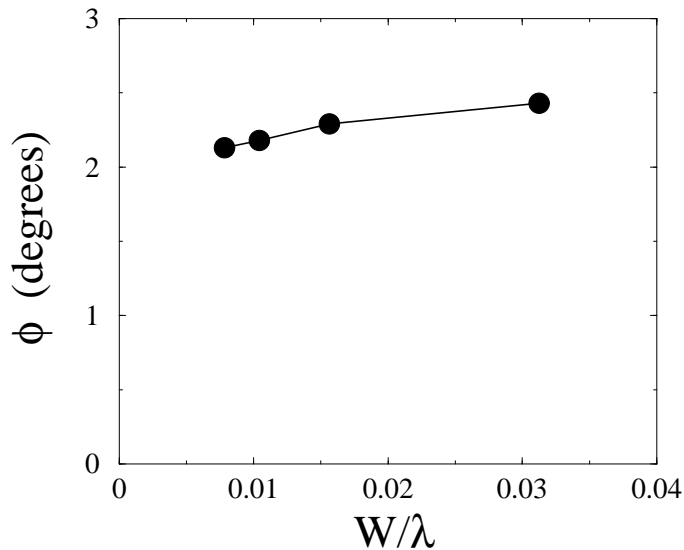,width=.5\textwidth}}
\caption{Trijunction rotation angle $\phi$ (see text for 
definition) versus $W/\lambda$ for the same series
of runs as Fig.~\protect\ref{figcbr}.
}
\label{figrot}
\end{figure}

Again, the phase-field result is fairly close to the boundary
integral calculation. However, this time a clearly visible
difference remains: Whereas the phase-field result is {\em above}
the boundary integral for the $\alpha$ lamella, it is {\em below}
for the $\beta$ lamella. Therefore, the difference is not a
simple shift between the two solutions. An explanation for this
fact comes from the detailed examination of the angles at the
trijunction point. Contrary to the completely symmetric situation
where $\alpha$ and $\beta$ are equivalent, the solid--solid interface
does not remain straight up to the trijunction. Rather, when the
trijunction is approached this interface starts to tilt and exhibits
a small angle $\phi$ with the vertical axis. The angles between
interfaces remain consistent with Young's law, but the 
whole trijunction zone is slightly rotated. Since we are in a 
steady state, the trijunction necessarily moves vertically 
(in the lab frame) as solidification proceeds. This implies
that the solid--solid interface is not the mere trijunction
trajectory, but must move after the trijunction passage, i.e.,
at its solid side, which is possible in a phase-field model, since the
diffusivity decays to zero in the solid over a distance 
of order $W$. In the sharp-interface formulation, in contrast,
where the diffusivity sharply falls
to zero in the solid, the solid--solid interface
cannot move and has to be strictly vertical up to the
trijunction. This difference between phase-field model and
sharp-interface description induces the observed shape difference
between the converged phase-field solution and the boundary-integral 
result. We plot in Fig.~\ref{figrot} the rotation 
angle $\phi$ as a function of $W/\lambda$. Contrary
to the deviation of the contact angle from its equilibrium value
found in Fig.~\ref{figangle}, this {\em global} trijunction rotation
angle does {\em not} extrapolate
to zero for $W/\lambda\to 0$, but rather seems 
to approach a finite value. This behavior will be discussed in more 
detail below.

\subsection{Oscillatory limit cycles}

As it is well known, lamellar steady-state growth becomes unstable
above a critical spacing that depends on the growth conditions
and the alloy composition. At the eutectic composition, the first
instability to occur is a period-preserving oscillatory instability,
that is, all lamellae of the same phase start to oscillate in phase. This
instability can be captured by our reduced simulation geometry,
because the center lines of each lamella remain symmetry planes
of the pattern, whereas the time translation symmetry of steady-state
growth is broken. The oscillations amplify in the course of time
and finally saturate and reach a limit cycle with a well-defined
amplitude. The bifurcation diagram of final oscillation amplitude
versus lamellar spacing was found to be slightly supercritical
in Ref.~\cite{Karma96}. Simulating oscillatory limit 
cycles close to the instability threshold therefore constitutes
a particularly sensitive test of our model, because even slight
deviations in the lamellar spacing or differences in behavior will induce 
large changes in the oscillation amplitude.

\begin{figure}
\centerline{\psfig{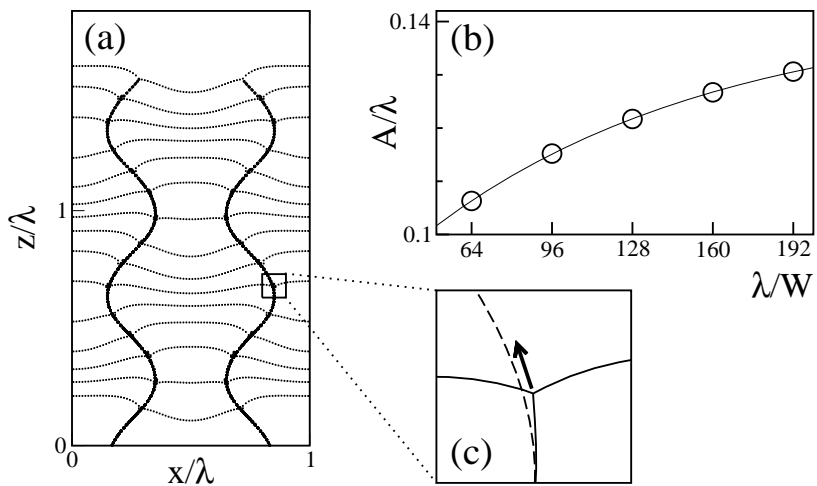}}
\caption{Oscillatory limit
cycle in the symmetric model alloy for 
$\lambda=2.2\, \lambda_{{\rm JH}}$. 
(a) Snapshots of the two solid--solid interfaces
(thick lines) and successive solid--liquid interfaces (thin lines).
(b): Oscillation amplitude $A/\lambda$ vs
inverse interface thickness, $\lambda/W$. (c): Blowup
of the trijunction region. The arrow in (c) indicates
the direction of the trijunction motion. 
}
\label{figshort}
\end{figure}

We used the symmetric model alloy with the same physical
parameters as in the preceding subsection, 
except the lamellar spacing, which was fixed to 
to $\lambda=2.2\, \lambda_{{\rm JH}}$. The simulations were started with
a slightly asymmetric initial condition in which one of the
lamellae was slightly ahead of the other. The limit
cycle rapidly emerged; an example for the simulated structure
once it has reached its final oscillation amplitude
is shown in Fig.~\ref{figshort}a for $\lambda/W=64$. 
We define the amplitude
$A$ of the limit cycle as the maximum lateral deviation
of the trijunction from its steady-state value, normalized
by the lamellar spacing, $A=\Delta x_{\rm max}/\lambda$.
We plot $A$ versus the resolution $\lambda/W$ in 
Fig.~\ref{figshort}b. It can be seen that the convergence 
is slow: the results still depend on the resolution
even for $\lambda/W=192$. 

One possible reason for this slow convergence can be recognized 
in Fig.~\ref{figshort}c, where we plot both a snapshot of a 
trijunction region (solid curves) and the final solid--solid interface left 
behind (dashed curve). Like for the asymmetric steady-state solutions, 
the trijunction seems to be slightly rotated, that is, 
the direction of the instantaneous solid--solid interface 
is not parallel to the direction of motion of the trijunction 
point (indicated in Fig.~\ref{figshort}c by an arrow
and roughly parallel to the final solid--solid interface).

\begin{figure}
\centerline{\psfig{file=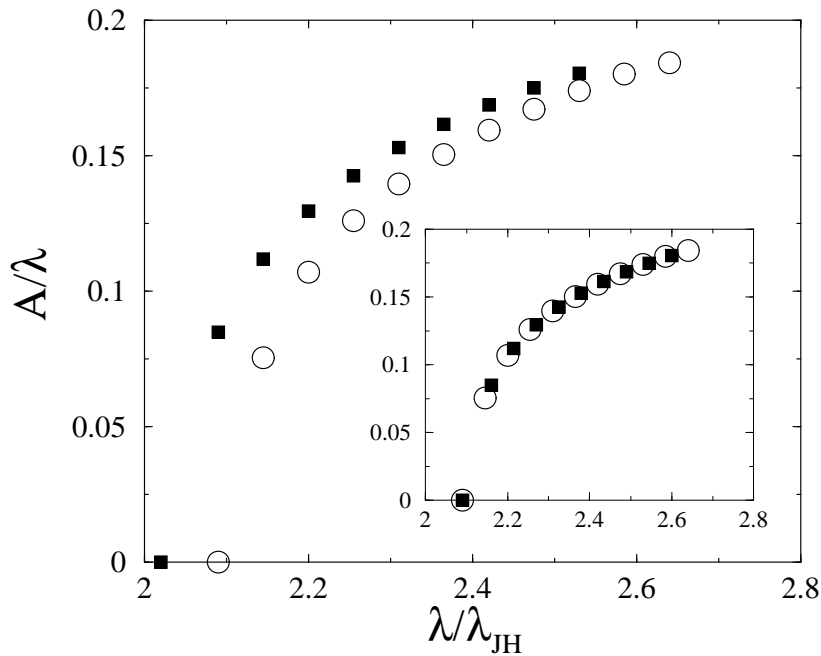,width=.6\textwidth}}
\caption{Amplitude of saturated oscillatory limit cycles
versus lamellar spacing, for parameters corresponding to
the first (circles) and third (squares) lowest resolution in
Fig.~\protect\ref{figshort}b ($\lambda/W=64$ and 128 there, respectively). 
Inset: same data, where
the squares have been shifted along the $\lambda$ axis
by $0.07$.
}
\label{figbifurc}
\end{figure}

Figure~\ref{figshort} seems to imply that our phase-field
model commits large errors when calculating limit cycles.
However, as already mentioned, we have chosen here a 
particularly sensitive test case. In Fig.~\ref{figbifurc}, 
we show the whole bifurcation diagram, that is, the amplitude 
of saturated limit cycles versus the lamellar spacing, calculated 
with two different resolutions. As shown in the insets, the
two curves can be superimposed quite nicely by a simple shift
along the $\lambda$ axis. Therefore, the relatively large difference in 
oscillation amplitudes comes from a small shift in the instability
threshold, while the overall shape of the bifurcation diagram
is quite robust. The relative error committed on the instability
threshold is about 3 \%. Hence, our model remains a useful tool 
for the exploration of nonlinear behavior in eutectic solidification,
even if the trijunction behavior prevents completely converged
calculations.

\subsection{Discussion}
\label{discustrijunction}

The most interesting outcome of the above numerical tests is the
information about the behavior of trijunction points. As already
mentioned, to our knowledge no detailed rigorous treatment of the 
equilibrium or dynamics of a diffuse trijunction is currently 
available, and it seems quite challenging indeed.
Therefore, only a qualitative explanation of our numerical findings 
can be given, namely by comparing the phase-field and sharp-interface 
pictures.

In the sharp-interface model, Young's law is assumed to hold
even outside of equilibrium; furthermore, the solid--solid interface
cannot move, since the solute diffusivity vanishes in the
solid immediately adjacent to the trijunction. In the phase-field
simulations, both of these assumptions are relaxed. As for
Young's law, there is a deviation from the equilibrium contact angles. 
In our simulations of the symmetric 
model alloy, this deviation decreases with the
interface thickness and extrapolates to zero in the (numerical)
sharp-interface limit. A likely origin of this effect is the 
relaxational time scale $\tau$ of the phase fields: A certain time
is necessary to establish local equilibrium against the ``pull'' 
of the advancing isotherms.

A more surprising result is the rotation of the trijunction as a whole,
which appears as soon as the symmetry between the two solid phases is broken. 
Here, we have shown only results at
the eutectic composition for an asymmetric phase diagram;
however, the
same effect is recovered in the symmetric model alloy at off-eutectic
composition, i.e., for unequal volume fractions of the two solids.
The rotation does not seem to vanish when the interface width tends 
to zero. Although we cannot, at present, give a detailed explanation
of this phenomenon, two speculative ideas about its origin might
be pursued in the future. The first is that this effect could 
result from the local equilibrium of curved diffuse interfaces:
Two different curvatures on the two sides of the trijunction could 
lead to some ``torque'' on the scale of the diffuse trijunction.
The second is that the solute diffusion field in the liquid ``wedge''
close to the trijunction yields a torque term {\em via} its
coupling to the diffuse interfaces, which would
be a purely non-equilibrium effect.

\begin{figure}
\centerline{\psfig{file=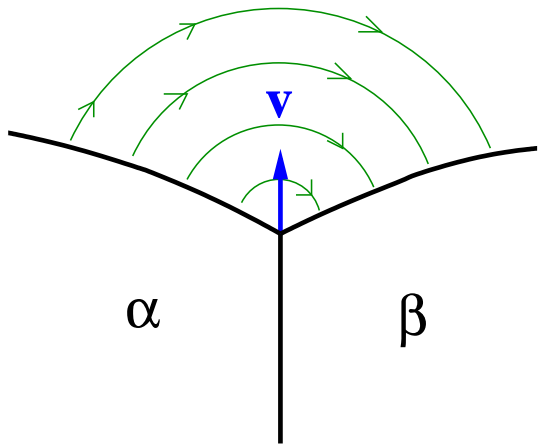,width=.4\textwidth}}
\caption{Sketch of the diffusion field around a trijunction
in the sharp-interface picture. Lines with arrows indicate
the flux of solute around the trijunction. Sufficiently
close to the trijunction, the concentration field obeys
the Laplace equation, which is known to have a singularity
in a wedge geometry. 
}
\label{figtrisketch}
\end{figure}

The origin of corrections to the classic sharp-interface model
that do not vanish for sharp interfaces can be qualitatively
understood. Contrary to the situation around a dendrite or cell
tip, where the lower cutoff scale for the diffusion field is
set by the local radius of curvature, a trijunction point is
a singular point in the sharp-interface solution. Indeed, when
the flux lines of the chemical components are drawn as sketched
in Fig.~\ref{figtrisketch}, it can be seen that the curvature of
these lines diverges as the trijunction point is approached.
Therefore, the sharp-interface solution is, in some sense,
inconsistent, because it is implicitly
based on an assumption of scale separation between interface
and ``macroscopic'' scales. Physical interfaces are smooth, 
and hence the local diffusion field
around a trijunction varies {\em on the same scale as the
interface thickness}. Therefore, the appearance of effects like
the trijunction rotation is, in itself, not surprising.
However, understanding them more quantitatively remains a
challenging problem. More simulations are needed to study in
detail the dependence of this phenomenon on the various
parameters such as growth speed, contact angles, phase diagram
and phase fractions.

Recently, the dynamics of trijunctions was found to have a 
noticeable influence on the stability of lamellar arrays at 
small spacings \cite{threshold}. More precisely, the
stability threshold for lamella elimination was found to
be distinctly smaller than the minimum undercooling spacing,
contrary to previous predictions \cite{jh,Langer80,Datye81}.
The reason is that the assumption used in these predictions
that trijunctions move in the direction normal 
to the large-scale envelope of the composite front (Cahn's rule)
was not satisfied.
It should be stressed that this
effect is not necessarily related to the trijunction
behavior found here, since Cahn's rule does not make any
assumptions on the {\em local} configuration around a
trijunction.

\section{Conclusions}
\label{conclusions}

We have presented and tested in detail a new phase-field
model for two-phase solidification. The specific design
of our free-energy functional has allowed us to clearly
separate, at a local level, 
the dynamics of interfaces from those of trijunctions,
and to eliminate all thin-interface corrections to the desired free-boundary
dynamics for each solid--liquid interface. 
Both properties have been checked by numerical simulations:
By the first we mean that interfaces between any two phases 
are completely free of the third phase far enough from the trijunctions,
and this is indeed verified both at equilibrium and for moving interfaces 
(see Fig.~\ref{figpfields}).
As for the second, it implies that simulations at fixed physical parameters 
become independent of the interface thickness, provided that it is 
not too large. To our knowledge, it is the first time
that the quantitative convergence to a
sharp-interface solution as the interface thickness is decreased, 
as shown in Fig.~\ref{figsymm},
has been explicitly demonstrated for a phase-field model
of solidification with more than one solid phase.

For generic, non-symmetric situations (unequal volume
fractions or physical properties of the two solid-liquid
equilibria), small differences exist between the
converged phase-field results and boundary-integral
simulations. Furthermore, while the convergence is very
satisfying for steady-state solutions, time-dependent
morphologies such as oscillatory limit cycles still
exhibit thin-interface effects. 
Whereas these prevent us, for the
moment, from achieving calculations that are completely
independent of the computational parameters, we have demonstrated
that the resulting errors are very small, except possibly
in the vicinity of bifurcation points. Since our model is
not only very accurate, but also efficient,
it is a suitable tool for the investigation of pattern
formation in two-phase solidification. In particular,
it can be applied to simulate three-dimensional structures
and to study their morphological stability \cite{Kurzproceedings}.

Because the individual interface dynamics
are perfectly distinct and controlled in our model, the only possible
source of the observed differences to sharp-interface models lies in
the trijunction region. Indeed, our detailed numerical
investigation has revealed that the behavior of the diffuse
trijunctions in the phase-field model differs slightly from
the assumptions usually made in sharp-interface theories,
as already discussed in Sec.~\ref{discustrijunction}.
Therefore, a quite surprising outcome of this study is that
the calculation of two-phase microstructures is far more
sensitive to the structure and dynamics of the
trijunction points than one might have expected. Note that these
findings have been possible only thanks to the specific
design of our model, which has allowed us to eliminate
all thin-interface corrections from the dynamics of the
solid-liquid interfaces, and to separate them from those of
the trijunctions. Therefore, our model is a privileged
tool to explore in detail the physics of diffuse trijunctions,
an interesting but challenging task. One conclusion
of our study is that, apart from its undeniable fundamental
interest, this more detailed knowledge is also a
necessary ingredient for a future quantitative modeling of
solidification with an arbitrary number of phases.

Let us conclude by sketching how our model can be extended to treat
more general situations. Here,
we have limited ourselves to eutectic diagrams with straight 
liquidus and solidus lines and constant concentration 
jumps; however, the structure of our model makes it possible 
to simulate any phase diagram. In particular, phase 
diagrams with curved coexistence lines as well as peritectic 
phase diagrams can be simulated without difficulty 
by choosing appropriate coefficients $A_i(T)$ and $B_i(T)$.

Next, because of our specific choice for the coupling between 
composition and phase fields, the model lacks the effect of 
the local interface curvature on the solute rejection. This
(negligibly small) effect could be recovered by the introduction 
of more complicated interpolation functions, as shown for 
single-phase solidification in Ref.~\cite{onesided}.

Furthermore, the asymptotic analysis and the antitrapping current in 
its present form are only applicable if the two solid--liquid
surface tensions are equal. In order to keep the benefits of the
model for arbitrary surface tensions, the asymptotic analysis
should be repeated (in part numerically) along the lines of
Ref.~\cite{onesided} as discussed in Sec.~\ref{unequal}; 
this is a straightforward albeit tedious calculation.

We expect that crystalline anisotropy can be incorporated in
the model in the standard way by making the interface thickness 
depend on the local orientation of the interfaces. However, 
care has to be taken in oder to ensure that the anisotropic
model still has interfaces that are free of third-phase
adsorption. Finally, some considerations on how to generalize
the model to more than three phases are presented in 
appendix \ref{morethanthree}.

\begin{acknowledgments}
We thank S. Akamatsu, G. Faivre, and A. Karma for many useful
discussions, and A. Karma for supplying us with the boundary 
integral code. This research was supported by the Centre National
d'\'Etudes Spatiales (France). R.F. was supported by 
the European Community through the Marie Curie program.
\end{acknowledgments}

\appendix

\section{Time constants}
\label{timeconstants}
Here, we show that a linear relation
between the time derivatives of the phase fields 
and the driving forces does not permit
to choose different relaxation times for different
interfaces and still guarantee that each interface 
is free of any third phase.

\begin{figure}
\centerline{\psfig{file=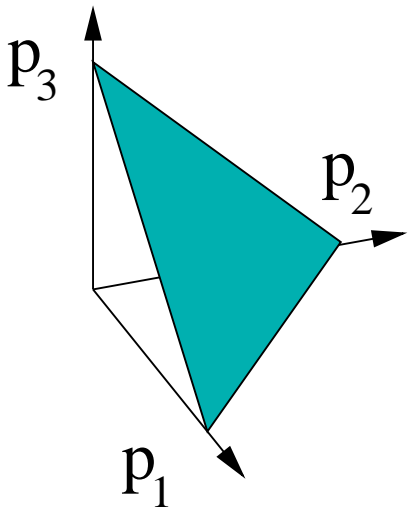,width=.3\textwidth}}
\caption{Sketch of the phase-field state space. The representative
point of the system is confined to the plane 
$p_1+p_2+p3=1$, whose intersection with the axes 
defines the vertices of the shadowed triangle, 
the so-called Gibbs simplex.
}
\label{figapp}
\end{figure}

A given configuration of the phase fields can be
visualized by a representative point in the 
three-dimensional vector space spanned by the
axes $p_1$, $p_2$ and $p_3$ (see figure \ref{figapp}).
The constraint $p_1+p_2+p_3=1$ 
restricts the allowed positions to a 
two-dimensional subspace, the plane that contains 
the points $(1,0,0)$, $(0,1,0)$, and $(0,0,1)$. These points
are also the vertices of an equilateral triangle, 
the Gibbs simplex (shadowed area). 
Only points inside the triangle (shadowed area) 
are meaningful if each phase field is to be interpreted
as a {\em positive} volume fraction.

The most general form for a linear relationship
between the time derivatives of the phase fields
and the driving forces $\delta F/\delta p_i$ is
\begin{equation}
\frac{\partial p_i}{\partial t} = 
  - \sum_{j=1}^3 \Gamma_{ij} \frac{\delta F}{\delta p_j},
\end{equation}
where $\Gamma_{ij}$ can be seen as a linear map between 
the vector of the driving forces, $\delta F/\delta p_i$,
and the vector of the time derivatives, 
$\partial p_i/\partial t$.
Since the representative point must remain in the
plane which contains the Gibbs simplex, 
the vector of the time derivatives 
must also lie in it, and therefore be normal 
to the vector $(1,1,1)$. In contrast, there are
no explicit constraints on the vector of the driving
forces, which can have any direction.
Therefore, the matrix $\Gamma_{ij}$ must have
the eigenvector $(1,1,1)$ with eigenvalue zero, or, 
in other words, be a projection on the allowed plane.

In addition, the free-energy density $F$ is designed
so that on purely binary interfaces there is no
driving force for the formation of the third
phase. For example, on the 1-2-interface, the
driving force along the $p_3$-direction is
zero; therefore, the vector  $\delta F/\delta p_i$
is parallel to $(1,-1,0)$. Since we do not want
the resulting time evolution to introduce
the third phase, the vector of the time derivatives must
also be parallel to $(1,-1,0)$. This implies that
this vector must be an eigenvector of the matrix
$\Gamma_{ij}$. The corresponding eigenvalue
is the relaxation rate for the 1-2-interface.

The same reasoning can be repeated for the 1-3-interface,
with the associated eigenvector $(1,0,-1)$.
Since the two vectors are not orthogonal, the
only possibility that makes both of them eigenvectors
is that the whole subspace corresponding to the plane of the
Gibbs simplex is an eigenspace with a
single eigenvalue. Therefore, all the time constants
associated with purely binary interfaces are the same, and
the only valid choice for the matrix $\Gamma$ is
a multiple of the projector 
on the plane of the simplex,
\begin{equation}
\Gamma_{ij}=\Gamma \left(\delta_{ij} - u_iu_j\right)
\end{equation}
where $\hat u=(1,1,1)/\sqrt{3}$ is the unit vector
normal to the simplex plane. Writing
$\Gamma = 1/(\tau H)$, the resulting equations of motion are
\begin{equation}
\tau \frac{\partial p_i}{\partial t} = 
   - \frac{1}{H} \sum_{j=1}^3 \left(\delta_{ij}-\frac{1}{3}\right) 
       \frac{\delta F}{\delta p_j}.
\end{equation}
This is equivalent to Eqs. (\ref{modelforp}) in conjuction
with the prescription of Eq. (\ref{prescription}).

The only way to achieve different relaxation times for
the different interfaces while keeping purely
binary interfaces is therefore to introduce
some nonlinearity in the relation between time derivatives
and driving forces. We achieve this by letting 
the time constant $\tau$ depend on the phase fields. Note 
that locally ``anisotropic'' dynamics in the configuration
space could also be induced by letting all the 
coefficients $\Gamma_{ij}$ depend on the phase fields,
so that the directions of the eigenvectors and the
associated eigenvalues depend on the location in the
configuration space.

\section{Symmetry of the gradient energy coefficients}
\label{generalgradients}

The gradient energy expression,
Eq. (\ref{fgrad}), satisfies the conditions for no
third-phase adsorption, Eqs. (\ref{valleys}).
We now consider its simplest generalization,
\begin{equation}
f_{\rm grad}=\sum_{mn} \frac{\xi_{mn}}{2} \vec\nabla p_m\cdot \vec\nabla p_n,
\end{equation}
a quadratic form in the gradients of the phase fields,
where the $\xi_{mn}$ are dimensionless constants 
that we assume to be symmetric (and positive), $\xi_{mn}=\xi_{nm}(>0)\;\forall m,n$.
We find that Eq. (\ref{flatness}) is satisfied if and only if the coefficients
that multiply the Laplacians of the phase fields in each purely binary interface
are all the same, so that this generalization does not provide any extra freedom
for our purposes. 

To see this, compute the functional, constrained derivatives of
a free-energy functional $F_{\rm grad}$ defined as the volume integral
of the gradient terms alone, $F_{\rm grad}=\int_V K f_{\rm grad}$. 
Making use of Eqs. (\ref{prescription})
and (\ref{funcderiv}), we find
\be
\label{rulewij}
-\left. \frac{\delta{F}_{\rm grad}}{\delta p_k}
     \right|_{p_\alpha + p_\beta + p_{\rm L}= 1} = K \left(
\sum_m \xi_{mk} \vec\nabla^2 p_m - \frac{1}{3} \sum_{mn} \xi_{mn} \vec\nabla^2 p_m
 \right).
\ee
The condition of flatness with respect to variations of $p_k$
on a $i$--$j$ interface, Eq.~(\ref{flatness}), then yields
\be
\label{wcond}
\xi_{ii} - 2 \xi_{ik} = \xi_{jj} - 2 \xi_{jk},
\ee
where we have taken into account that $p_k\equiv 0$ and $p_j=1-p_i$.  
This can be written in a more symmetric way by adding $\xi_{kk}$
to both sides:
\be
\xi_{kk} - 2 \xi_{ik} + \xi_{ii} = \xi_{kk} - 2 \xi_{jk} + \xi_{jj}.
\ee
Two equivalent conditions can be obtained for the other
two interfaces, so that all three symmetric combinations
$\xi_{mm} - 2 \xi_{mn} + \xi_{nn}$ must have the same value.
 
Next, replace $k$ with $i$ in Eq.~(\ref{rulewij}) and,
afterwards, apply it to a $i$--$j$ interface ($p_k=0$).
One finds
\be
-\left. \frac{\delta{F}_{\rm grad}}{\delta p_i}
     \right|_{p_\alpha + p_\beta + p_{\rm L}= 1} =
\frac{K}{3} ( 2 \xi_{ii} + \xi_{jj} - 3 \xi_{ij} - \xi_{ik} + \xi_{jk})
\nabla^2 p_i.
\ee
Using Eq.~(\ref{wcond}) to eliminate $\xi_{jk}-\xi_{ik}$, 
we find that the prefactor of the
Laplacian can be rewritten as $(K/2)(\xi_{ii} - 2 \xi_{ij} + \xi_{jj})$,
i.e., one half of the combination that we saw above to be the same
for all interfaces. Therefore, we can choose $\xi_{ij}=\delta_{ij}$
without loss of generality, and we are back to
the minimal model with Eq. (\ref{fgrad}).

The multi-phase-field models based on the approach of 
Steinbach {\it et al.} \cite{Steinbach96} use a different expression for the 
gradient terms. It combines gradients of the phase fields
with the fields itself, and is hence not a simple quadratic 
form in the gradients. We have checked that this expression 
does not satisfy the condition of flatness; therefore, it is 
not a suitable alternative for our purposes.

\section{More than three phases}
\label{morethanthree}
In principle, the multi-phase formalism where each phase 
field represents a local volume fraction can be extended 
to an arbitrary number of phases. However, the methods
we have applied to ensure that interfaces are free of any
adsorbed third phase need to be generalized with care.
Let us consider a multi-well potential $f_{\rm MW}$ constructed
in the same manner as our triple-well potential $f_{\rm TW}$, i.e.,
by a superposition of individual double-well potentials 
for each phase field, $f_{\rm MW}=\sum_i f_{\rm DW}(p_i)$. 
For $f_{\rm DW}(p)=|p|^v|1-p|^v$ (with $v$ a positive real) 
the free energy of a point where $n$ phases
coexist ($n$-junction) is $n^{1-2v}(n-1)^v$. For our model, $v=2$, 
this expression has a single maximum at $n=3$. 
Therefore, a quadrijunction already has a lower free
energy than a trijunction, which means that the free-energy
landscape $f_{\rm MW}$ has undesirable local minima 
for more than three phases. 
These lower-energy $n$-junctions can always be ``raised'' 
by adding one obstacle term $f_{\rm obs}=\Pi_{i=1}^n p_i^2$ per 
$n$-junction to treat. However, the number of such terms
becomes rapidly prohibitive when $n$ increases; in addition,
such ``steep'' terms require a higher numerical resolution.
Alternatively, for $0<v\le1$ the $n$-junction
energy becomes monotonously increasing in $n$ for all positive $n$.
Therefore, the simplest choice ensuring the right topography
for an arbitrary number of phases would be $v=1$,
a model considered in Ref. \cite{interface}; however,
to our knowledge no thin-interface analysis has been performed 
to date for this inverse parabolic potential.


\begin{thebibliography}{99}
\bibitem{foot3} This is the main difference to the first version
  of our model published in \protect\cite{dresden}.
\bibitem{footonehalf}
The same evolution equations on an $i$--$j$ interface can be obtained by
taking the functional derivative of the free energy evaluated on that 
interface with respect to either of the non-vanishing phase fields. 
This is possible thanks to the fact that the derivatives normal 
to that interface vanish, so that the operators corresponding to setting 
$p_k=0$ and taking the functional derivative commute. 
However, by this method, since one eliminates $p_j$ in terms of $p_i$ 
before taking the derivative with
respect to $p_i$, the contribution of $p_i$ is double-counted,
so that the result needs to be divided by two.
\bibitem{foot2} The values of the constants $a_1$ and $a_2$ are related
  to those calculated by Karma and Rappel in Ref.~\protect\cite{kr} and 
  denoted here by $a_1^{KR}$ and $a_2^{KR}$, by $a_1=(8/15)a_1^{KR}$ and 
  $a_2=(15/8)a_2^{KR}$.
\bibitem{foot1} Note that the partition relation, Eq.~(\ref{partition}), 
  has to be replaced by $C_-=C_+ - (C_{\rm L}-C_i)$ when $k_i\to 1$.
\end{thebibliography}
\end{document}